\documentclass[sigconf,nonacm]{acmart}

\usepackage{bbding}
\usepackage{pifont}
\usepackage{wasysym}

\AtBeginDocument{%
  }

\newcommand{\name}{UB-Mesh}

\newcommand{\costsaving}{$2.04$}

\title{\name: a Hierarchically Localized nD-FullMesh Datacenter Network Architecture}
\usepackage{makecell}
\usepackage{multirow} 
\usepackage[normalem]{ulem}
\usepackage[linesnumbered, ruled, vlined]{algorithm2e}
\usepackage{pifont}
\usepackage{amsfonts}
\usepackage{framed}
\usepackage{graphicx}

\usepackage{titlesec}
\makeatletter
\g@addto@macro{\normalsize}{%
  \setlength{\abovedisplayskip}{3pt plus 0.5pt minus 1pt}
  \setlength{\belowdisplayskip}{3pt plus 0.5pt minus 1pt}
  \setlength{\abovedisplayshortskip}{0pt}
  \setlength{\belowdisplayshortskip}{0pt}
  \setlength{\intextsep}{4pt plus 1pt minus 1pt}
  \setlength{\textfloatsep}{4pt plus 1pt minus 1pt}
  \setlength{\skip\footins}{5pt plus 1pt minus 1pt}
  }
\makeatother

\titlespacing\subsection{0pt}{2pt plus 1pt minus 1pt}{3pt plus 1pt minus 2pt}
\titlespacing\subsubsection{0pt}{2pt plus 1pt minus 1pt}{3pt plus 1pt minus 2pt}

\usepackage{enumitem}

\author{Heng Liao\textsuperscript{\Envelope}, Bingyang Liu\textsuperscript{\Envelope}, Xianping Chen, Zhigang Guo, Chuanning Cheng, Jianbing Wang, \\ Xiangyu Chen, Peng Dong, Rui Meng, Wenjie Liu, Zhe Zhou\textsuperscript{*}, Ziyang Zhang, Yuhang Gai, \\ Cunle Qian, Yi Xiong, Zhongwu Cheng, Jing Xia, Yuli Ma, Xi Chen, Wenhua Du, Shizhong Xiao, Chungang Li, Yong Qin,Liudong Xiong, Zhou Yu, Lv Chen, Lei Chen, Buyun Wang, Pei Wu, \\ Junen Gao,Xiaochu Li, Jian He, Shizhuan Yan, Bill McCOLL \\ \emph{Huawei}}

\begin{document}

\thispagestyle{plain}
\pagestyle{plain}
\begin{abstract}

    As the  Large-scale Language Models (LLMs) continue to scale, the requisite computational power and bandwidth  escalate. To address this, we introduce \name, a novel AI datacenter network architecture designed to enhance scalability, performance, cost-efficiency and availability.
    Unlike traditional datacenters that provide symmetrical node-to-node bandwidth, \name~employs a hierarchically   localized nD-FullMesh network topology. This design fully leverages the data locality of LLM training,  prioritizing short-range, direct interconnects 
    to minimize data movement distance and reduce switch usage.

    Although \name's nD-FullMesh topology offers several theoretical advantages, its concrete architecture design, physical implementation and networking system optimization present new challenges.  
    For the actual construction of \name, we first design the UB-Mesh-Pod architecture, which is based on a 4D-FullMesh topology. \name-Pod is implemented via a suite of  hardware components that serve as the foundational building blocks, including specifically-designed NPU, CPU, Low-Radix-Switch (LRS), High-Radix-Switch (HRS), NICs and others. These components are interconnected via  a  novel Unified Bus (UB) technique, which enables flexible IO bandwidth allocation and hardware resource pooling.  For networking system optimization,  we propose advanced routing mechanism named All-Path-Routing (APR) to efficiently manage data traffic. These optimizations, combined with topology-aware performance enhancements and robust reliability measures like 64+1 backup design, result in  \costsaving $\times$ higher cost-efficiency, 7.2\% higher network availability compared to traditional Clos architecture and 95\%+ linearity in various LLM training tasks.

\end{abstract}

\maketitle

\section{Introduction}

\newcommand\blfootnote[1]{%
\begingroup
\renewcommand\thefootnote{}\footnote{#1}%
\addtocounter{footnote}{-1}%
\endgroup
}

\blfootnote{\textsuperscript{\Envelope} Corresponding authors.}
\blfootnote{\textsuperscript{*} Drafted the paper.}
\label{sec:intro}
The emerging Large Language Models (LLMs) \cite{gpt4,llama31,ren2023pangu, llama, chatglm, Gemma2, bai2023qwentechnicalreport, yang2023baichuan2openlargescale, ai2024yiopenfoundationmodels, guo2024deepseekcoderlargelanguagemodel, deepseekai2024deepseekv2strongeconomicalefficient} are transforming the AI industry and human society. Following the Scaling Laws~\cite{kaplan2020scaling,clark2022unified}, LLMs continue to enhance their understanding, generation, and reasoning capabilities by expanding model parameters and training data. This trend, however, presents increasing challenges to the underlying training systems and infrastructures, forcing next-generation AI datacenters to meet the following requirements:
\begin{figure} [t]
    \centering
    \includegraphics[width=1.0\linewidth]{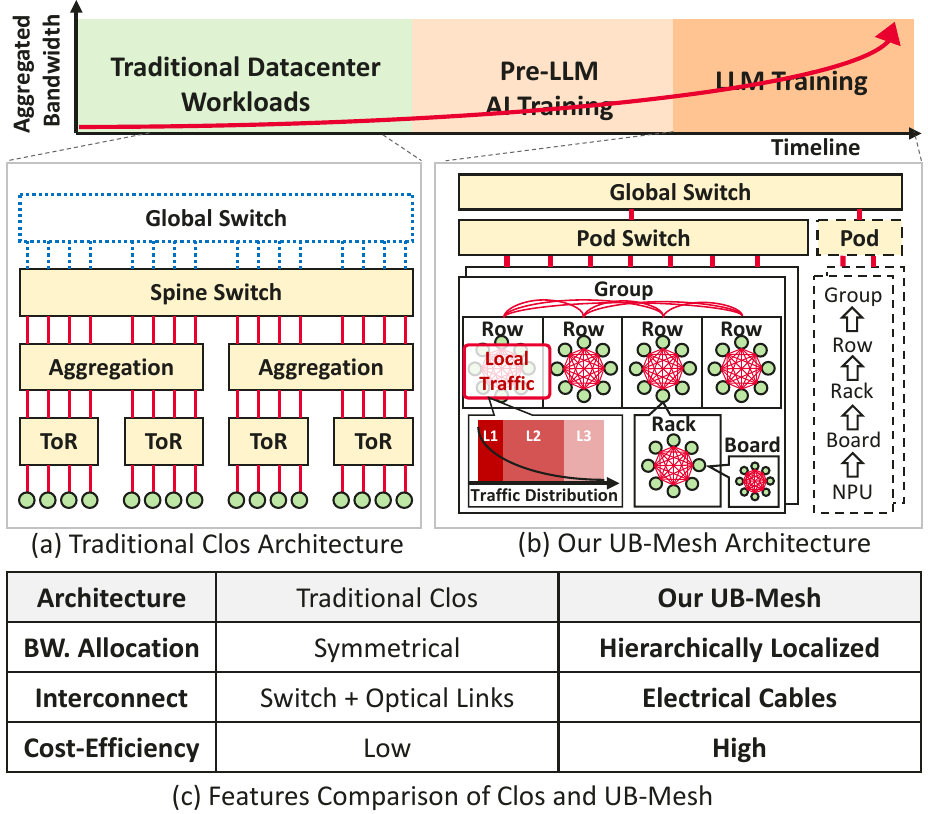} 
    \caption{{Comparison of Traditional Clos Datacenter Architecture and UB-Mesh}}
         \label{fig:dcn}
\end{figure}

\emph{\textbf{R1: Large Scale.}} As model size and training data volume escalate, an increasing number of NPUs (Neural Network Processing Units) or GPUs are required to complete training within a reasonable timeframe. For instance, the pretraining of LLAMA-3 requires 54 days with 16K GPUs~\cite{llama31}.  Recent announcements from leading companies have shown that AI training systems with  100K GPUs have been successfully deployed~\cite{xai_100k}.  Scalable  infrastructures are necessary for supporting the continuous evolution of LLM technologies.
 
\emph{\textbf{R2: High Bandwidth.}} In LLM training systems, AI compute nodes (NPU/GPU) require interconnect bandwidth exceeding 3.2 Tbps per node~\cite{nvidia2023dgx, gaudi3}, roughly 10 times greater than typical CPU node interconnects in contemporary data centers. As a result, the aggregate bandwidth of state-of-the-art AI training systems is 10 to 100 times higher than that of current CPU-based Infrastructure-as-a-Service (IaaS) systems.

\emph{\textbf{R3: Cost Efficient.}} Building a large-scale AI datacenter necessitates a substantial hardware investment, often reaching billions of dollars in capital expenditure (CapEx). To achieve the required 10 to 100$\times$ increase in aggregate interconnect bandwidth, using traditional symmetrical Clos data center network architecture (illustrated in Fig.~\ref{fig:dcn}-(a)) would result in interconnect costs also increasing 10 to 100$\times$. Optimizing network infrastructure presents a significant opportunity to enhance cost efficiency. Furthermore, reducing the operational costs (OpEx), including energy consumption and maintenance, are equally critical to ensure overall cost-effectiveness.

\emph{\textbf{R4: High Availability.}} A large-scale LLM training cluster with 100K computation nodes and roughly 1 million optical modules faces significant availability challenges. Current statistics indicate that even with a Mean Time Between Failures (MTBF) of 5 years per link, the raw MTBF of the whole 100K-GPU AI cluster drops to less than 30 minutes~\cite{semianalysis2023}. To address this, the network architecture design must not only improve hardware reliability but also incorporate fault-tolerance mechanisms  to handle failures in interconnects, compute resources, control systems, and storage.

Achieving these objectives simultaneously is extremely challenging and requires a paradigm shift towards advanced datacenter network architecture design. We argue that three principles should be central to designing next-generation AI datacenters:

\emph{\textbf{P1: Traffic-Pattern Driven Network Topology.}} 
Unlike traditional datacenter workloads, which typically generate uniform and random data traffic, large language model (LLM) training traffic is deterministic and exhibits strong data locality. For example, the collective communication required by \emph{Tensor Parallelism}~\cite{shoeybi2019megatron} operations often accounts for over 50\% of the total data traffic and typically occurs within a cluster of 8-64 adjacent NPUs. In contrast, collective communication arising from \emph{Data Parallelism} accounts for less than 2\% of total traffic, which however usually requires long-range transmission.
Therefore, a hierarchical, localized network architecture is essential to align with these traffic patterns.

\emph{\textbf{P2: Topology-Aware Computation\&Communication.}} Given a hierarchical  localized datacenter network, effectively running LLM training on them presents another significant challenge. If training tasks are not optimally distributed across computing resources or if the networking system is not well-optimized, AI clusters may suffer from low performance due to traffic congestion or bandwidth  under-utilization. To address this, the choice of parallelism strategies, routing, collective communication, load balancing, and other factors must be meticulously aligned with the network topology.

\emph{\textbf{P3: Self-Healing System for Fault-Tolerance.}} The LLM training system must incorporate self-healing capabilities to ensure robustness. In the event of link failures, the routing system should automatically switch to an alternate path. Similarly, if an NPU fails, there should be a mechanism  to seamlessly activate a backup NPU, maintaining system integrity and LLM training continuity.

To meet the requirements \emph{\textbf{R1-R4}} and adhere to the principles \emph{\textbf{P1-P3}}, we introduce an innovative \name~architecture. As illustrated in Fig.~\ref{fig:dcn}-(b), \name~employs a nD-FullMesh network topology, which recursively constructs full-mesh topologies—beginning with 1-D connections between adjacent NPUs on a board, expanding to 2-D connections between adjacent 1-D meshes within a rack, and scaling to 3-D and beyond by interconnecting adjacent higher-dimensional meshes across broader ranges. This hierarchically localized network architecture minimizes transmission hops and optimizes per-hop distance. It prioritizes direct interconnects over indirect, long-range switching to reduce the reliance on switches and optical modules, thereby fulfilling \emph{\textbf{R1}} and \emph{\textbf{R3}}. Additionally, it enables hierarchical bandwidth allocation based on the transmission requirements of LLM training, providing high bandwidth for short-range communication and lower bandwidth for long-range communication, which meets \emph{\textbf{R2}} and {\textbf{R3}} and adheres to \emph{\textbf{P1}}.

Adhering to \emph{\textbf{P2}}, we delve into advanced networking and system optimization mechanisms to enhance the \name~architecture. Specifically, we introduce the  All-Path Routing (APR) technique to fully utilize the bandwidth of direct-connection links. APR combines Source Routing, Structured  Addressing\&Linear Table Lookup and  Deadlock-Free Flow Control mechanisms to enable adaptive routing, minimize forwarding overhead and avoid deadlock. Additionally, we incorporate topology-aware fast-fault recovery to  enhance reliability. To further optimize performance, we propose topology-aware collective communication and parallelization optimizations to improve bandwidth utilization during training.

To meet \textbf{\emph{R4}} and adhere to \emph{\textbf{P3}}, \name~employs a 64+1 high-availability design: each rack consists of an additional backup NPU. In the event of unexpected failures in the NPUs within the system, the backup NPU is activated to restore functionality and ensure the uninterrupted continuation of LLM training jobs. Additionally, the routing system facilitates rapid failure recovery in the case of link failures through a novel direct-notification technique.

 We meticulously engineer the hardware and system stack of \name, considering various engineering constraints and tradeoffs. Our concrete implementation, the \name-Pod~\footnote{In this paper, we use \emph~{UB-Mesh-Pod} to denote the concrete implementation of the \name~architecture, with its nD-FullMesh topology dimension set to 4 to balance flexibility and cost-efficiency.}, features a 4D-FullMesh topology. This architecture enables \name~to scale seamlessly up to 8K NPUs, forming a high-bandwidth domain capable of supporting the construction of next-generation AI datacenters. To achieve this, we have developed a suite of hardware components that serve as the foundational building blocks, including the NPU, CPU, Low-Radix-Switch (LRS), High-Radix-Switch (HRS), NICs and others.
 
 Note that, in contrast to baseline systems that employ diverse interconnect techniques (e.g., PCIe, NVLINK, IB and RoCE), \name~utilizes a novel \textbf{Unified-Bus (UB)} technique for all component interconnections. This unified approach improves the flexibility of IO resource allocation and its peer-to-peer communication capability enables efficient hardware resource pooling. UB also provides opportunities for seamless cross-layer optimization.

 Comprehensive evaluations demonstrate that, compared to a non-oversubscribed Clos network,  \name~reduces high-radix switch usage by 98\% and optical module usage by 93\%, achieving a \costsaving $\times$ improvement in system-level cost-efficiency. Experiments across multiple LLM training tasks also show that UB-Mesh achieves marginal performance degradation (within 7\%) compared to a costly  Clos network. This combination of low cost and high performance  not only meets the current demands of LLM training but also positions it to address future scalability challenges effectively.

\section{Background \& Motivation}
\label{sec:background}

\subsection{"Communication Wall" in LLM Training}

LLM training is the largest scale and most computation and communication intensive parallel computing application ever~\cite{shoeybi2019megatron,huang2019gpipe, ren2023pangu, llama, chatglm, deepspeed, ds-moe, gpt4}. Obeying the so-called scaling laws~\cite{kaplan2020scaling, clark2022unified, firefly}, the performance of LLM models is improved via scaling-up the model parameters and training data. Consequently, LLMs are demanding increasing numbers of AI accelerators to finish training within a reasonable time. For example,   the open-sourced LLAMA-3.1 models are trained on 16K GPUs~\cite{llama31}, and the next-generation LLM models are using 100K GPUs for training~\cite{semianalysis2023, wccftech_2023}. 

A standard training procedure entails the repetition of training iterations, where each iteration comprises a forward pass to calculate losses, a backward pass to determine gradients, and an optimizer step to adjust the model parameters. To fully utilize distributed computation power, LLM training splits the data, model and activations to  tens of thousands of NPUs via various  parallelism strategies.  Within each iteration, the NPUs frequently exchange data with each other  to distribute input data,  synchronize activations and gradients, etc.  As the training system scales,  data movement becomes the most expensive part of the system~\cite{hpn, gangidi2024rdma}. Without strong inter-NPU communication capabilities, the training process is easily bottle-necked by the "Communication Wall".

\subsection{Locality of Data Traffic in LLM Training}

\label{sec:locality}

\begin{figure} [t]
    \centering
    \includegraphics[width=0.98\linewidth]{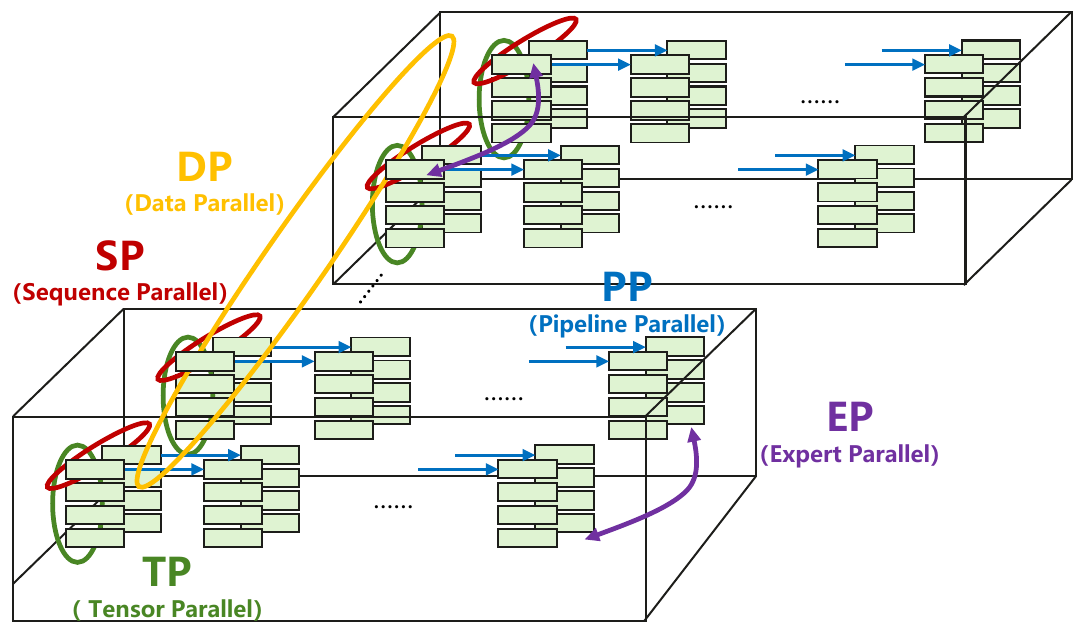} 
    \caption{{ Parallelism  in LLM Training}}
         \label{fig:llm_training_pattern}
\end{figure}

As illustrated in Fig.\ref{fig:llm_training_pattern}, LLM training typically involves multiple parallelism techniques, which are introduced as follows: 

\textbf{Tensor Parallelism: } TP splits model layers in a  row-wise or column-wise manner and places sub-layers on multiple NPUs that compute in parallel~\cite{shoeybi2019megatron}. It mainly involves \texttt{AllReduce} operations to merge the distributed partial results. 

\textbf{Sequence Parallelism: }  SP (also referred to as Context Parallelism in some papers) is usually adopted to split sequences to multiple NPUs to realize parallel processing. SP relies on Ring-Attention~\cite{liu2023ring} techniques and can employ \texttt{AllGather} operations to gather partial results in different NPUs.

\begin{table}[t]
    \caption{Analysis of Data Traffic in LLM Training}
    \label{tab:ai_dataflow}
    \resizebox{0.48\textwidth}{!}{\renewcommand{\arraystretch}{1.2}{
        \begin{tabular}{c|ccccc}
            \toprule
            \textbf{\begin{tabular}[c]{@{}c@{}}Parallism\\ Techniques\end{tabular}} & \textbf{\begin{tabular}[c]{@{}c@{}}Communication \\ Pattern\end{tabular}} & \textbf{\begin{tabular}[c]{@{}c@{}}Data Volume \\ Per Transfer\end{tabular}} & \multicolumn{1}{l}{\textbf{Total Transfer}} & \multicolumn{1}{l}{\textbf{Total Volume}} & \multicolumn{1}{l}{\textbf{Data Traffic}} \\  
            \midrule
            TP                                                                & AllReduce                                                                 & 360 MB                                                                       & 4992                                        & 1775 GB                                   & 52.9\%                                    \\
            SP                                                                & AllGather                                                            & 180/360 MB                                                                   & 4992/1664                                   & 1462.5 GB                                 & 44.08\%                                   \\
            EP                                                                & AlltoAll                                                                  & 10.5 MB                                                                      & 4992                                        & 51.19 GB                                  & 1.54\%                                    \\
            PP                                                                & P2P                                                                       & 192 MB                                                                       & 26                                          & 4.875 GB                                  & 0.14\%                                    \\
            DP                                                                & AllReduce                                                                 & 711.75 MB                                                                    & 64                                          & 44.48 GB                                  & 1.34\% \\
            \bottomrule                                  
            \end{tabular}
    }}
    \end{table}
\textbf{Expert Parallelism: } For LLM models adopting Mix-of-Experts (MoE) techniques, dense MLP layers are replaced by MoE layers, where each layer contains several "Experts". Experts are sparsely activated during execution.  EP distributes experts on different NPUs. The input tokens are dynamically sent to target experts  via \texttt{All2All}  communication.

\textbf{Pipeline Parallelism: } Unlike TP that splits each model layer, PP distributes layers to multiple devices, and execute forwards and backwards in a pipelined manner. PP involves low-overhead \texttt{P2P} communication for transmitting activations across layers, but requires efficient scheduling algorithms~\cite{narayanan2019pipedream, huang2019gpipe}  to  minimize  bubbles.

\textbf{Data Parallelism: } DP replicates models and optimizer states across multiple NPUs. Each replica processes a portion of input batches in parallel. \texttt{AllReduce} operations are necessary to synchronize gradients during the training procedure.

These parallelism techniques collectively distribute training tasks across thousands of NPUs.
It is important to note that \textbf{\emph{not all parallelism techniques generate equal volumes of data traffic.}} According to our analysis based on an in-house MoE-2T model, as detailed in Table~\ref{tab:ai_dataflow}, communication intensity is hierarchical and exhibits strong locality. Specifically, TP and SP account for approximately 97\% of the total traffic, while the remaining parallelism techniques typically generate less than 2\% of the total traffic. Other model architectures may exhibit slightly different data traffic distributions but also demonstrate strong locality~\cite{wangrail}. Therefore, \textbf{\emph{the architecture design should prioritize a hierarchical approach to network bandwidth provisioning.}}

\begin{figure} [t]
    \centering
    \includegraphics[width=0.99\linewidth]{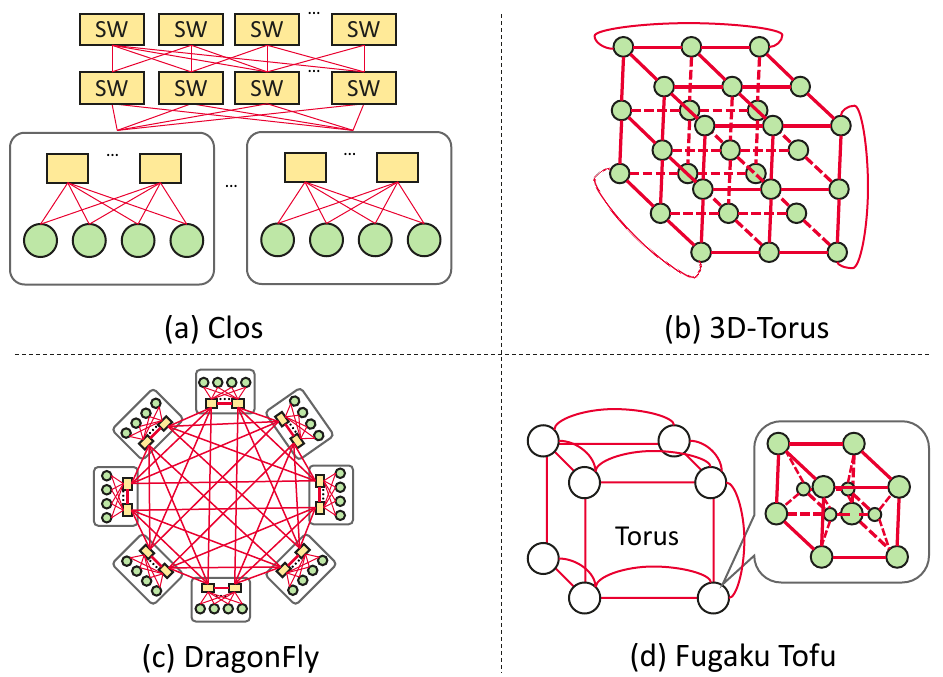} 
    \caption{{Traditional Datacenter Network Architectures}}
         \label{fig:baseline_topo}
\end{figure}

\subsection{Datacenter Network Architectures}

Designing a large-scale LLM training cluster involves one of the most critical considerations: how to organize numerous NPU resources into a cohesive distributed system? This involves determining the appropriate datacenter network architecture. Traditional data centers and supercomputing systems have explored lots of topologies, but these may not be well-suited for large-scale LLM training. Fig.~\ref{fig:baseline_topo} illustrates some popular topologies:

\textbf{Clos}:
  Traditional datacenters usually adopt the Clos architecture, which employs two or three tiers of switches to symmetrically connect NPUs/CPUs, offering high performance and adaptability to various traffic patterns. However, its high cost is a drawback, primarily due to the extensive use of high-performance switches and optical modules.

\textbf{3D Torus}:
Unlike  Clos, which relies on switch-based interconnects, the 3D Torus connects adjacent NPUs directly. 
While reducing costs, the 3D Torus offers lower NPU-to-NPU bandwidth and is less accommodating to complex communication patterns such as All-to-All communication, which is prevalent in MoE models~\cite{ds-moe}.

\textbf{Dragon-Fly}: The Dragon-Fly topology~\cite{Dragonfly} organizes switches into interconnected groups, where each switch within a group is directly connected to every other switch. These groups are further linked via long-range cables, allowing packets to be routed indirectly. This architecture offers reduced dimensionality in traditional datacenter and HPC environments. DF is cheaper than Clos but still costly due to full NPU-switch bandwidth requirements, and its architecture performs poorly for LLM training traffic, especially in P2P/AllReduce scenarios.

\textbf{Fugaku Tofu}: Fugaku Tofu~\cite{ajima2009tofu} proposes a unique 6D Torus topology for HPC applications. However,  similar to 3D Torus, it offers lower NPU-to-NPU bandwidth and is less accommodating to complex communication patterns.

In summary, how to design a high performance and cost-efficient datacenter network architecture for large-scale LLM training is still an open problem.

\section{\name~Architecture}

To meet requirements \textbf{\emph{R1-R4}} and adhere to principles \textbf{\emph{P1-P3}}  in Section~\ref{sec:intro}, we propose \name~architecture. \name~ adopts a novel nD-FullMesh topology that maximizes the use of direct electrical-cable interconnects, thereby reducing the need for expensive high-bandwidth switches and optical modules. We then present a series of hardware modules specifically designed to construct \name. Finally, we will discuss the detailed architecture design of the \name-Pod, the concret 4D-FullMesh implementation of \name, with several engineering constraints and tradeoffs fully considered.

\subsection{The nD-FullMesh Topology}

\begin{figure} [t]
    \centering
    \includegraphics[width=1.0\linewidth]{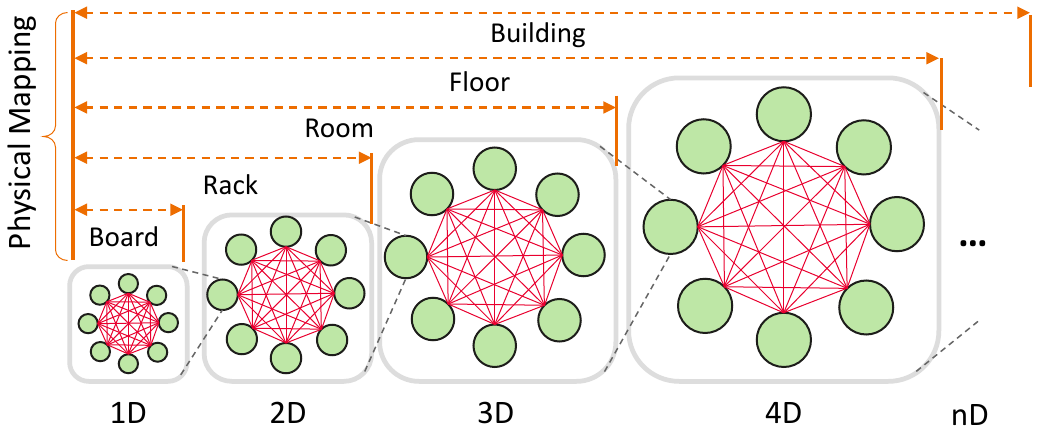} 
    \caption{{\name's nD-FullMesh Topology and Possible Physical Mapping}}
         \label{fig:ndfullmesh}
\end{figure}

As illustrated in Fig.~\ref{fig:ndfullmesh}, \name~introduces a novel nD-FullMesh topology that emphasizes direct interconnects across all network tiers. This topology begins with adjacent nodes forming a 1-D full-mesh, where each node is connected to every other node in the same tier. Extending this concept, nodes within adjacent 1-D full-meshes are then directly interconnected, creating a 2-D full-mesh. This process continues recursively, with adjacent 2-D full-meshes forming a 3-D full-mesh, and so on, ultimately scaling to an n-D full-mesh as needed. This flexible virtual topology can be seamlessly mapped to various physical NPU organizations: 1-D full-meshes on individual boards, 2-D full-meshes within racks, 3-D full-meshes across rack rows, 4-D full-meshes within a floor's rack groups, and even 5-D full-meshes spanning entire buildings, and beyond.

\begin{figure} [t]
    \centering
    \includegraphics[width=1.0\linewidth]{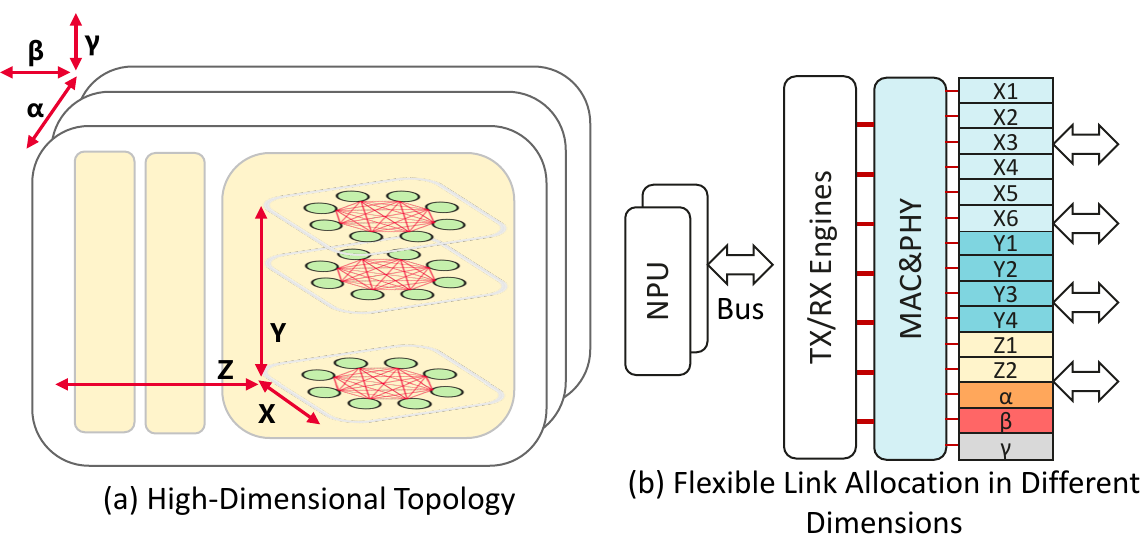} 
    \caption{{High-Dimensional Topology and Flexible Bandwidth Allocation\protect\footnotemark }}
         \label{fig:link_allocation}
       
\end{figure}

Such a hierarchically localized topology has several advantages in building next-generation LLM training datacenters. First, since nD-FullMesh forms tightly-coupled direct-connection domain in each network dimension and can provide tier-over-tier oversubscribed bandwidth,  it fully  leverages the data locality and dense-to-sparse traffic patterns inherent in LLM training (Sec.\ref{sec:locality}). Also, we can flexibly adjust the per-node bandwidth allocation for each dimension to meet the specific requirements from future LLM training tasks. For instance, as shown in Fig.~\ref{fig:link_allocation}-(a), assuming that there is a 6D-FullMesh topology, we can adjust the bandwidth  in $XYZ$ and $\alpha\beta \gamma$ dimensions by allocating different number of interconnect links in NPUs' IO modules, as illustrated in Fig.~\ref{fig:link_allocation}-(b).   

Secondly, unlike Clos-based architectures that demand extensive switching, this direct-connected topology provides great potential of reducing transmission distance. Combined with data placement and collective communication optimization (Sec.\ref{sec:cc}), most transmissions (typically involving \emph{TP} or \emph{SP} as per Table~\ref{tab:ai_dataflow}) can be completed in 0-2 hops, substantially minimizing the data movement overhead.

Thirdly, compared to other popular topologies like Clos, the nD-FullMesh topology mitigates the reliance on high-bandwidth switches and long-range optical interconnections. Instead, it maximizes the use of short-range direct interconnects. For example, as estimated in Table~\ref{tab:link_choice}, short-range passive electrical cables account for 86.7\% of the total consumed cables according to our estimation. This design substantially reduces the cost of switches and optical modules, while also enhancing system reliability, as electrical cables and connectors are more stable than optical modules. 
\footnotetext{The number of ports and their allocation in the figure is just for illustration rather than real implementation.}

\begin{table}[t]
    \caption{Usage Estimation of Different Types of Links }
    \label{tab:link_choice}
    \resizebox{0.48\textwidth}{!}{\renewcommand{\arraystretch}{1.0}{
  \begin{tabular}{cccr}
    \toprule
\textbf{Dimension} & \textbf{Distance}               & \textbf{Link Type}       & \textbf{Ratio}     \\
$X Y$              &  $\sim1$ m & Passive Electrical Cable   &   86.7\%     \\
$Z$                &  $\sim10$ m & Active Electrical Cable   &    7.2\%  \\
$\alpha$           &  $\sim10^2$ m & Optical Cable          &     4.8\%   \\
$\beta \gamma$  ...            &  $\sim10^3$ m  & Optical Cable        &      1.2\%   \\
\bottomrule
\end{tabular}
    }}
    \end{table}

\begin{table*}[t]
    \centering
    \caption{Main Building Blocks of \name.}
    \small
    \label{tab:building_blocks}
    \setlength{\tabcolsep}{2.5mm}{
    \renewcommand{\arraystretch}{1}{
    \resizebox{1.0\textwidth}{!}{
    \begin{tabular}{m{1.7cm}<{\centering}|m{4.1cm}<{\centering}|m{3.1cm}<{\centering}|m{3.1cm}<{\centering}|m{3.1cm}<{\centering}}
    \toprule
                Hardware         & \textbf{NPU} & \textbf{CPU} & \textbf{LRS} & \textbf{HRS} \\ \midrule
    Illustration       &       $\includegraphics[height=0.6in]{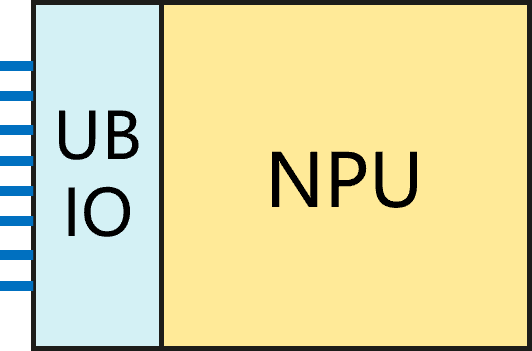}$     &    $\includegraphics[height=0.6in]{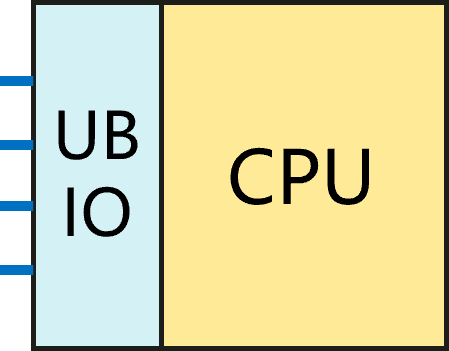}$      &     $\includegraphics[height=0.6in]{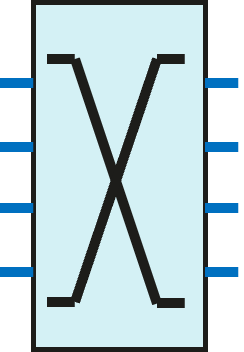}$          &   $\includegraphics[height=0.6in]{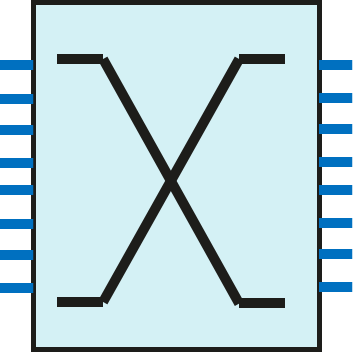}$    \\ \midrule
    Functions & AI compute    &   Host CPU  &  Low-radix switching     &  High-radix switching    \\
    \midrule
    IO Capability & UB   x72    &    UB x32     &   UB x72         &UB x512  \\
     \bottomrule
    \end{tabular}} } 
    }
\end{table*}
Lastly, unlike the 3D-Torus topology, which is not well-suited for complex collective communication operations like \texttt{all-to-all}, the full-mesh topology can efficiently support these operations, which will be discussed in Sec.~\ref{sec:system_perf_opt}.

\subsection{Building Blocks of \name}

The proposed nD-FullMesh topology offers several theoretical advantages, but its practical implementation and the enhancement of system performance present new challenges, necessitating a comprehensive consideration of various engineering constraints and tradeoffs. To address these, we have designed and manufactured a series of hardware modules that serve as the fundamental building blocks of \name. These modules are interconnected through a novel interconnection mechanism called \textbf{Unified Bus (UB)}. The details are as follows:

\subsubsection{\textbf{Hardware Modules.}}

Table~\ref{tab:building_blocks} enumerates the primary components of \name. At its core, the NPU serves as the AI compute unit. The NPU equips two UB IO controllers,  which collectively provide x72 UB lanes. The CPU, responsible for executing host programs, is also equipped with one UB IO controller, offering a UB x32 IO. 

Although \name~prioritizes direct interconnects, switching capabilities remain essential to meet specific requirements. For example, it is better to aggregate the $Z$ and $\alpha\beta\gamma$-dimension NPU IOs in Fig.~\ref{fig:link_allocation} to facilitate efficient inter-rack interconnection. Additionally, we expect that the CPU-NPU ratios and their binding relationships can be flexibly adjusted to realize efficient NPU and memory resource pooling, which can be realized via switching.
 
To provide cost-efficient switching capabilities within \name-Pod, we develop \emph{LRS}, which are lightweight, low-radix switches. It has low manufacturing cost and efficiently aggregates inter-rack IO bandwidth and enables CPU-NPU communication.  

Despite that \name~can be implemented using low-radix switches alone, practical network constraints often limit its scope to a certain size (\name-Pod). Therefore, \name~also includes high-radix switches (HRS) that offer a UB x512 IO capacity for Pod-level switching. Detailed architecture design and considerations of \name~are introduced in Sec.~\ref{sec:arch}.

\subsubsection{\textbf{The Unified Bus (UB) Interconnects.}}
\begin{figure} [t]
    \centering
    \includegraphics[width=0.99\linewidth]{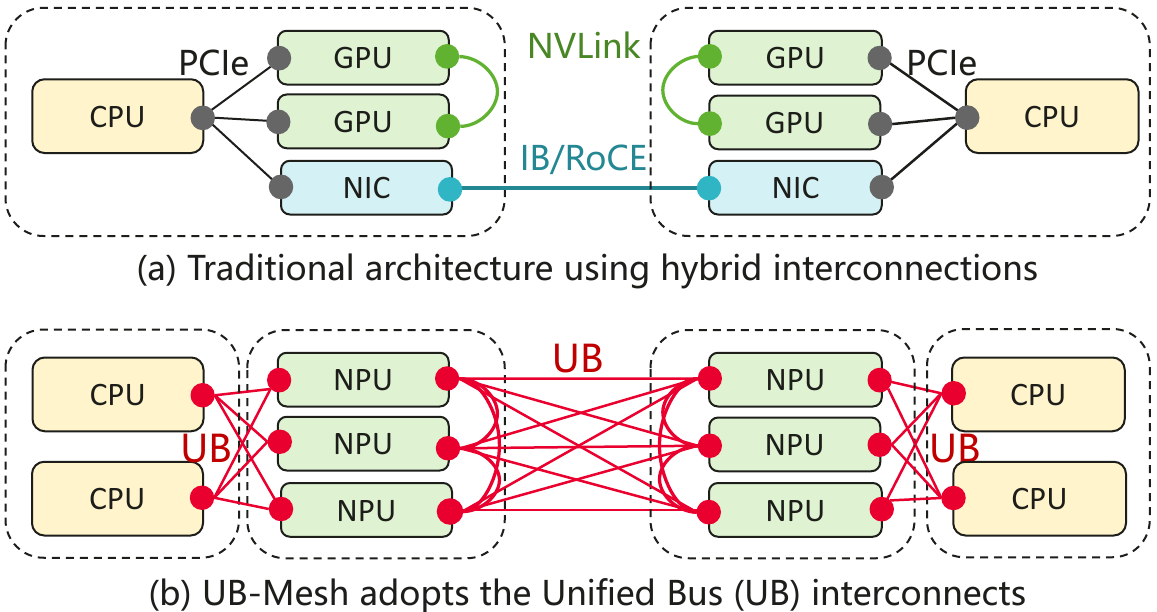} 
    \caption{{The Unified Bus Interconnects VS. Traditional Hybrid Interconnects}}
         \label{fig:ub}
\end{figure}

\begin{figure*} [t]
    \centering
    \includegraphics[width=0.99\linewidth]{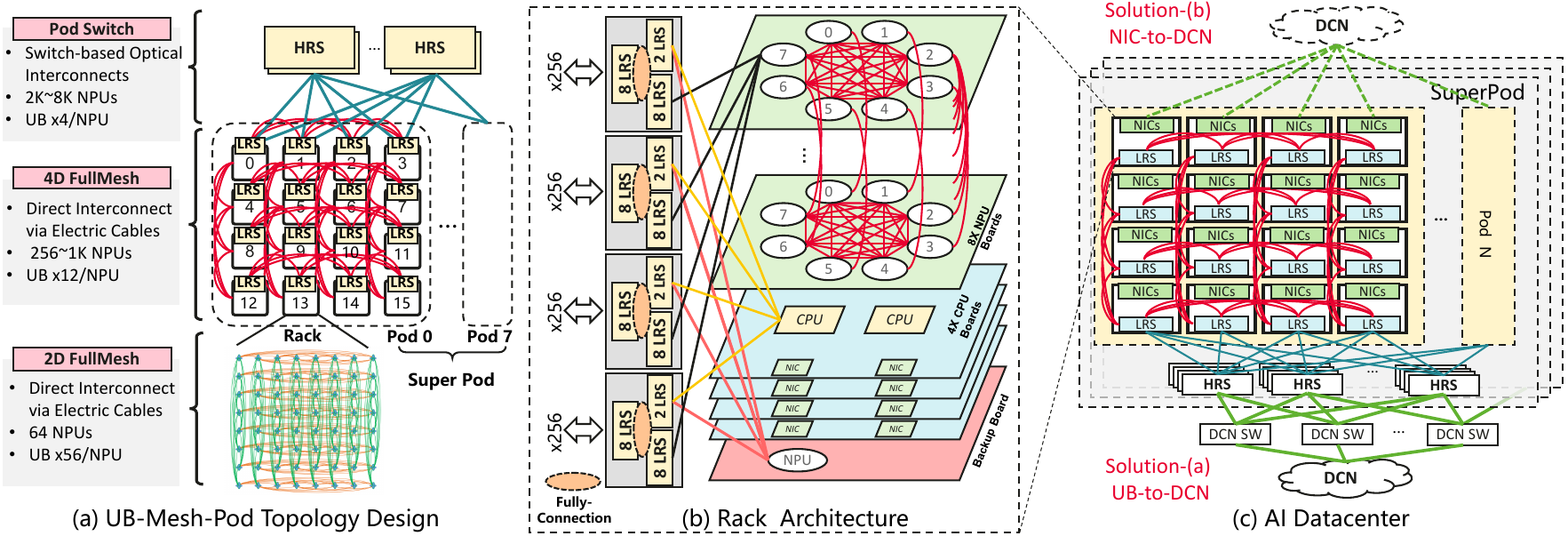} 
    \caption{{ Architecture Design of \name}}
         \label{fig:overview}
         \vspace{-0.5em}
\end{figure*}

\label{sec:ub_interconnects}
Connecting these heterogeneous hardware modules requires specific interconnect techniques. As illustrated in Fig.\ref{fig:ub}-(a), traditional GPU-based LLM training systems\cite{hpn, gangidi2024rdma} typically adopt hybrid interconnects: PCIe for CPU-GPU and CPU-NIC interconnection, NVLink for intra-server GPU interconnection, and InfiniBand/RoCE for inter-server interconnection. As shown in Fig.~\ref{fig:ub}-(b),  unlike previous designs, we introduce the Unified Bus (UB) interconnect\footnote{The detailed specification of UB protocol is going to be released.}.  UB simultaneously supports synchronized operations  like Load/Store/Atomic and asynchronous operations like Read/Write/Message. Based on UB, we only need to design and implement a single UB IO controller that can be reused across CPU, NPU, and even LRS switches.  UB also offers the following three main advantages:

\begin{itemize}[leftmargin=1em]
\item \textbf{Flexible IO Resource Allocation.} The UB interconnects are decoupled from specific use cases, enabling flexible resource allocation for different types of IO within the chip, as illustrated in Fig.~\ref{fig:link_allocation}. The inter-NPU bandwidth and CPU-NPU bandwidth can also be flexibly adjusted according to specific requirements, as they utilize the same UB links.
\item \textbf{Hardware Resource Pooling.} The UB's peer-to-peer communication capabilities enable efficient pooling of hardware resources, including DDR DRAM, CPUs, NPUs and NICs. For instance, in Fig.~\ref{fig:ub}-(b), CPUs and NPUs are pooled via UB interconnects to enhance resource utilization. 
\item \textbf{System Optimization Benefits.} A unified interconnect eliminates the need for protocol conversions, significantly reducing overhead and simplifying the design and optimization of drivers, communication libraries, and frameworks, among others. 
\end{itemize}

\subsection{Architecture Overview} 

\label{sec:arch}

Composing the hardware building blocks listed in Table~\ref{tab:building_blocks} and interconnecting them via the Unified Bus (UB), we present the overall architecture design of \name. As illustrated in Fig.~\ref{fig:overview}-(a), we implement a \name-Pod following the proposed nD-FullMesh topology. Specifically, within this architecture, we create a 2D-FullMesh within each rack and extend this to another 2D-FullMesh beyond the rack, resulting in a 4D-FullMesh. We have chosen not to scale \name~to 5D-FullMesh or higher in this generation to achieve an engineering balance between cost-efficiency and flexibility, still adopting a Clos topology beyond the 4D-FullMesh\footnote{\name-Pod~can evolve to 5D-FullMesh or higher dimensions as future LLM training demands scale.}. Details are introduced as follows:

\subsubsection{\textbf{Implementing 2D-FullMesh in Racks.}}

\begin{figure} [t]
    \centering
    \includegraphics[width=1.0\linewidth]{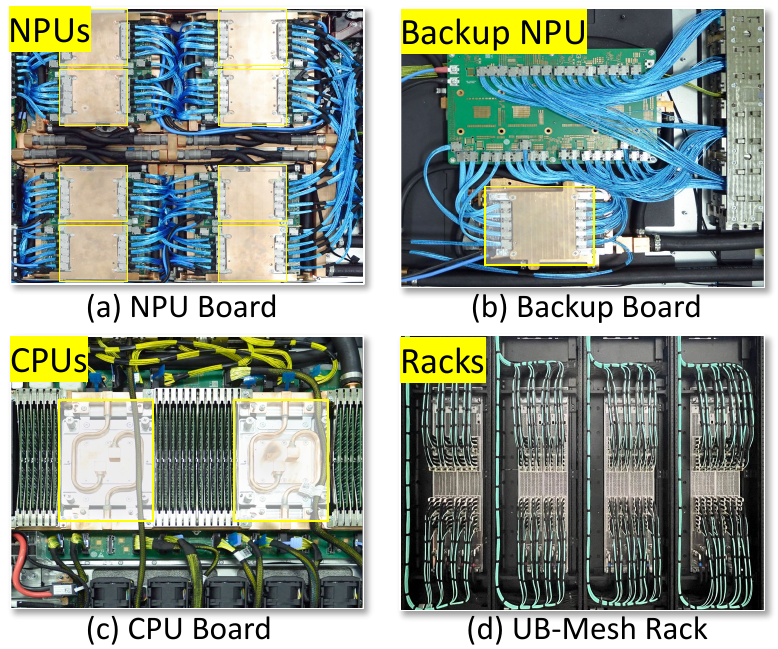} 
    \caption{{The Hardware Implementation of \name~Rack}}
         \label{fig:photo}
\end{figure}

As illustrated in Fig.~\ref{fig:overview}-(b), each rack is configured with a 2D-FullMesh topology, integrating multiple components to ensure efficient communication and resource utilization. The core of the rack consists of eight NPU boards, each containing eight NPUs, as shown in Fig.~\ref{fig:photo}-(a). These 64 NPUs are interconnected within the rack to form a 2D-full-mesh network, ensuring high NPU-to-NPU bandwidth. Note that since the UB IO controllers also have the routing ability, each NPU serves as a router and enables indirect routing in this architecture. 

In addition to the NPUs, the rack includes dedicated CPU boards, as shown in Fig.~\ref{fig:photo}-(c).   Unlike traditional setups where CPUs and NPUs are on the same board, here they are separated. The CPUs connect to the NPUs through switches, enabling flexible CPU/NPU ratios and supporting CPU/NPU/DDR resource pooling for resource utilization improvement.

The rack features multiple back-plane switches that manage both intra-rack and inter-rack connections. These switches employ low-radix designs, termed \emph{LRS}, to reduce costs while ensuring non-blocking communication among devices. As shown in Fig.~\ref{fig:overview}-(b), 18 LRSes are fully-connected to form one switch plane, where two LRSes are used for CPUs and backup NPUs, eight for regular NPUs and eight for inter-rack connection. Aggregately, these back-plane switches output four UB x256 IO.

\subsubsection{\textbf{The 64+1 High Availability Design.}}
As highlighted in Section~\ref{sec:intro}, ensuring system availability is a critical challenge in large-scale AI datacenters. Individual NPUs within the system may experience unexpected failures. While the software system can detect these failures and restart training using the remaining healthy NPUs, the system performance will be significantly impacted due to reduced computational power and system bandwidth.

\begin{figure} [t]
    \centering
    \includegraphics[width=0.85\linewidth]{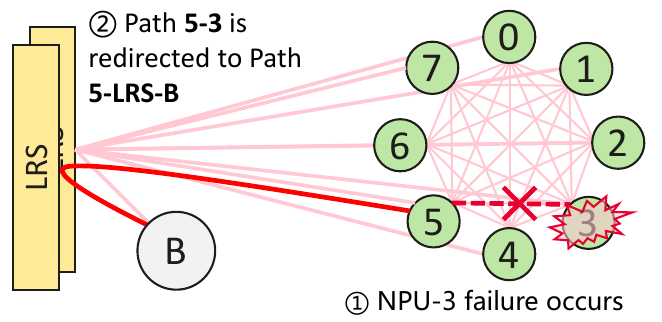} 
    \caption{{Fault-Tolerance via Enabling Backup NPU}}
         \label{fig:backup}
\end{figure}

To enhance system availability, we introduce a special 64+1 design: As depicted in Fig.\ref{fig:overview}-(b) and Fig.\ref{fig:photo}-(b), the system comprises 64 regular NPUs along with an additional backup NPU. This NPU is connected to the 64 regular NPUs via LRS. The backup NPU is strategically utilized in the event of any NPU failures. As illustrated in Fig.~\ref{fig:backup}, when NPU-3 has a failure, the management system activates the backup NPU (node-B in the figure) to replace NPU-3. The original direct-connection links related to NPU-3 are redirected. For instance, the path 5-3 is  redirected to path 5-LRS-B. Although this strategy changes the original direct-connection to one-hop routing, slightly increasing transmission latency, it is far superior to simply masking NPU-3 and running tasks on the remaining seven NPUs. In LLM training scenarios where intra-rack bandwidth is the primary concern, the increased latency has a negligible impact on overall training performance.

\subsubsection{\textbf{Scaling to 4D-FullMesh in \name-Pod.}}

Within a rack, we organize 64 NPUs in a 2D full-mesh topology. Beyond the rack, we implement another 2D full-mesh to organize the racks, forming a 4D-FullMesh architecture named \name-Pod.

As depicted in Fig.\ref{fig:overview}-(a) and (c), four adjacent racks in a row are organized in a tightly-coupled 1D full-mesh, where racks are directly interconnected via LRS ports in each rack. In each rack column, racks are also directly connected, creating a 16-rack 2D full-mesh topology. Each link in this configuration represents a UB x128 IO, as shown in Fig.\ref{fig:photo}-(d). We connect four adjacent racks along two dimensions for constructing inter-rack full-mesh because this is the optimal point considering the reach of active electrical cables. Since each rack has 64 NPUs and each pod has 16 racks, a 4D-FullMesh \name-Pod contains 1024 NPUs in total.

\subsubsection{\name-SuperPod and Beyond.}

Having the 1K \name-Pod, we further construct the \name-SuperPod, which can hold multiple \name-Pods. Considering that in current cloud scenarios the small or middle-scale LLM training workloads may not consume the whole SuperPod, we choose to adopt the symmetrical Clos topology in the Pod-level interconnection, rather than continue the usage of full-mesh. Such a design  allows the cloud managers to flexibly divide the SuperPod according to consumers' requirements and guarantee full bandwidth in each divided domains. 
As shown in Fig~\ref{fig:overview}-(c), we use high-radix Pod-switches (HRS) to connect each rack in the SuperPod, scaling  up to 8K NPUs. 

Finally, racks in SuperPods are also connected to the large-scale DCN (Data-Center Network) either via UB switches (\emph{Solution-(a)}) or via the NICs located on CPU boards (\emph{Solution-(b)}). The DCN domain usually supports large-scale ~\emph{Data Parallelism} training. The DCN switches are  organized using Clos topology. The DCN domain can scale to 100K NPUs or more.

\section{ Enabling The  Architecture}

\begin{table}[t]
    \caption{Routing Systems Comparison}
    \label{tab:routing}
    \setlength{\tabcolsep}{1mm}
    \resizebox{0.48\textwidth}{!}{\renewcommand{\arraystretch}{0.9}{
        \begin{tabular}{c|c|c|c|c}
            \toprule
            \begin{tabular}[c]{@{}c@{}}Routing \\ System\end{tabular}             & LPM with BGP                                                                              & \begin{tabular}[c]{@{}c@{}}Host-based \\ Routing\end{tabular}                           & DOR                                                                  & APR    \\ \midrule
            Application                                                           & Generic DCN                                                                               & IB                                                                                      & Tofu,TPU                                                             & UB-Mesh \\ \midrule
            \begin{tabular}[c]{@{}c@{}}Design\\ Principle\end{tabular}            & \begin{tabular}[c]{@{}c@{}}Shortest Path First\\ + Longest Prefix \\ Matching\end{tabular} & \begin{tabular}[c]{@{}c@{}}Host-based Routing\\ +EM/Linear Table \\ Lookup\end{tabular} & \begin{tabular}[c]{@{}c@{}}Dimension Ordered \\ Routing\end{tabular} & \begin{tabular}[c]{@{}c@{}}Topology-Aware\\ Routing\end{tabular}      \\ \midrule
            \begin{tabular}[c]{@{}c@{}}Support \\ Hybrid\\ Topology\end{tabular}  &          \textcolor{green}{\CheckmarkBold}                                                                                 &            \textcolor{green}{\CheckmarkBold}                                                                             &       \textcolor{red}{\XSolidBrush}                                                                &       \textcolor{green}{\CheckmarkBold}    \\ \midrule
            \begin{tabular}[c]{@{}c@{}}High-Performance\\ Forwarding\end{tabular} &           \textcolor{red}{\XSolidBrush}                                                                                 &            \textcolor{red}{\XSolidBrush}                                                                              &   \textcolor{green}{\CheckmarkBold}                                                                   &      \textcolor{green}{\CheckmarkBold}     \\ \midrule
            \begin{tabular}[c]{@{}c@{}}Non-Shortest\\ Path\end{tabular}           &          \textcolor{red}{\XSolidBrush}                                                                                  &         \textcolor{red}{\XSolidBrush}                                                                                 &          \textcolor{red}{\XSolidBrush}                                                             &       \textcolor{green}{\CheckmarkBold}    \\ \midrule
            \begin{tabular}[c]{@{}c@{}}Fault\\ Tolerance\end{tabular}             &         \textcolor{green}{\CheckmarkBold}                                                                                  &            \textcolor{green}{\CheckmarkBold}                                                                             &         \includegraphics[height=0.13in]{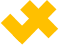}                                                            &     \textcolor{green}{\CheckmarkBold}      \\ \bottomrule
            \end{tabular}
    }}
\end{table}

In \name-SuperPod, the 4D-FullMesh + Clos topology urgently demands efficient hybrid routing support: apart from switch-based routing via LRS and HRS, the NPUs themselves also have routing capabilities via the UB controllers. The large number of routers and the hierarchical topology in \name~introduce new complexities to the routing system. We identify that the routing system must fulfill the following five key requirements:

\begin{itemize}[leftmargin=1em]
\item \textbf{Support for Hybrid Topology:}  The routing system must meet the requirements incurred by \name's 4D-FullMesh + Clos hybrid topology.

\item \textbf{Efficient Forwarding:} Since each NPU is also a router, and the entire system contains a large number of NPUs, to conserve NPU hardware resources, the routing system must efficiently handle routing-table lookups and forwarding operations.

\item \textbf{Support for Non-Shortest Paths:} In the nD-FullMesh topology, between two endpoints, there are several possible paths with diverse distances. The system should enable the use of non-shortest paths to maximize bandwidth utilization across the network.

\item \textbf{Rapid Failure Recovery:} To ensure reliability and availability, the routing system must quickly recover from any failures, reducing the impact on the training procedure.

\item \textbf{Deadlock Free:} Finally, the entire network system must operate without the risk of deadlock, ensuring smooth and uninterrupted data flow.
\end{itemize}

 However, existing routing techniques used on Clos, Torus and DragonFly topologies, like Longest-Prefix-Matching (LPM), Host-based Routing and Dimension Ordered Routing (DOR) as listed in Table~\ref{tab:routing}, none of them satisfy all our requirements. Therefore, we propose an All-Path-Routing (APR) technique and Fast Fault-Recovery via Direct Notification techniques to meet the requirements.

\subsection{All-Path Routing (APR)} 

\label{sec:apr}
\begin{figure} [t]
    \centering
    \includegraphics[width=0.92\linewidth]{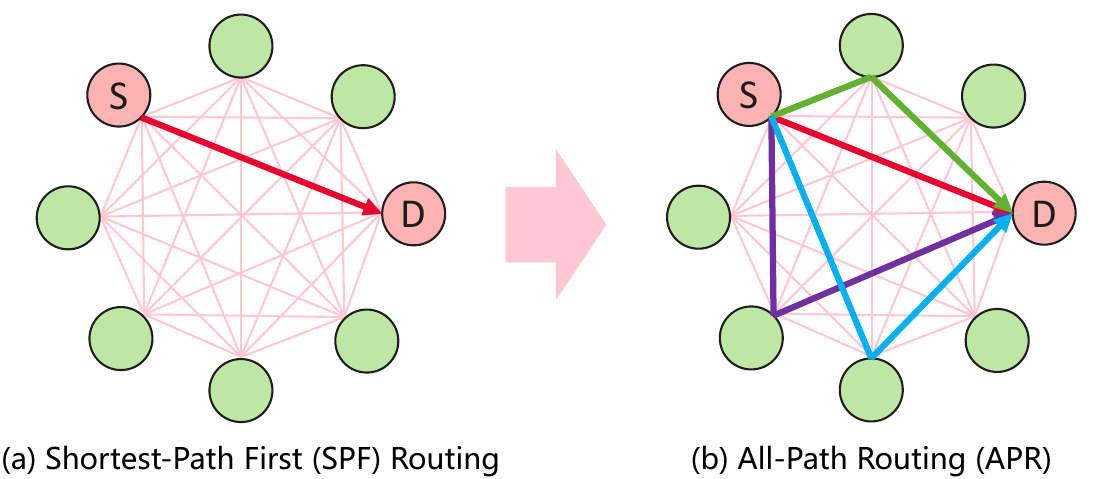} 
    \caption{{ Shortest-Path-Routing VS. All Path Routing }}
         \label{fig:indirect_routing}
\end{figure}

\name's full-mesh network architecture provides multiple paths between any two NPUs. Traditional routing strategies like Shortest-Path First routing (SPF), as illustrated in Fig.~\ref{fig:indirect_routing}-(a),  often underutilizes network bandwidth and is susceptible to link failures. To enhance network efficiency and resilience, we propose the All-Path Routing (APR) mechanism in \name.

As shown in Fig.~\ref{fig:indirect_routing}-(b), APR leverages all available paths between source and destination nodes.  
This flexibility allows for dynamic path switching, which responds to failures or congestion, thereby improving network robustness. To achieve this goal efficiently, APR leverages three underline mechanisms:(1)  \emph{Source Routing} (SR) mechanism to realize adaptive routing.  (2) \emph{Structured Addressing\&Linear Table Lookup}  technique for minimize the routing-table lookup and forwarding overheads.  (3) \emph{Topology-Aware Deadlock-free Flow Control} for deadlock avoidance.

\subsubsection{\textbf{Source Routing.}}
\begin{figure} [t]
    \centering
    \includegraphics[width=0.98\linewidth]{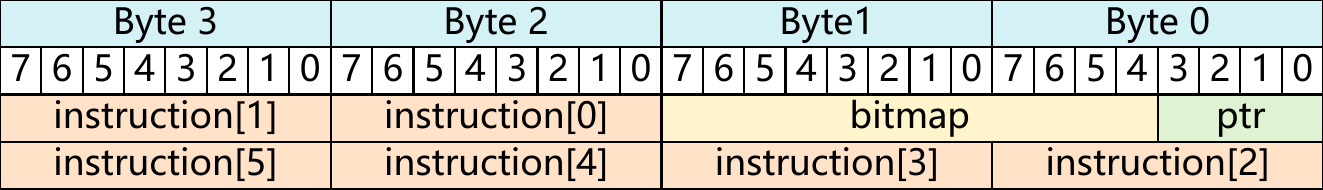} 
    \caption{{Source Routing Header Format}}
         \label{fig:sr}
\end{figure}

In order to fully utilize the paths provided by APR, one practical way is leveraging the Source Routing (SR) mechanism. As shown in Fig.~\ref{fig:sr}, an 8-byte SR header, containing the forwarding instructions, is added to original 
packet header. Each router uses a 4-bit \texttt{ptr} to indicate a bit offset in the 12-bit bitmap field. The value of the $i$-th bit specifies
the way of forwarding in the $i$-th hop (i.e., bit = 1 indicates SR forwarding in this hop, 0 for traditional forwarding). In case of SR forwarding, \texttt{Bitmap} field
is also used to locate one of the six  instruction fields, which
tells how to forward this packet. The SR information is highly compacted
and then of low overhead.

\subsubsection{\textbf{Structured Addressing\&Linear Table Lookup.}}

To minimize the routing-table lookup and forwarding overheads in each NPU's UB IO controller, the APR routing system leverages the structure of \name's topology and employs a structured addressing and linear table lookup mechanism. Specifically, the addressing space is divided into segments based on the physical location of network elements, such as Pods, racks, and boards. Since NPUs within a segment share the same prefix, only the short segment address needs to be stored, and NPUs can be addressed via linear offsets relative to the segment address. This design significantly reduces table space, expedites route table generation and distribution, accelerates convergence, and enables swift response to state changes, such as in failure recovery operations.

\subsubsection{\textbf{Topology-Aware Deadlock-free Flow Control.}}
Given that \name's nD-FullMesh topology contains complex ring structures and the APR mechanism also enables multi-hop routing, it is challenging to realize deadlock-free flow control with limited VL (Virtual Lane) resources. To overcome this challenge,  we propose the TFC (Topology-Aware Deadlock-Free Flow Control) algorithm. This algorithm minimizes VL resource usage while enabling deadlock-free all-path routing with only 2 VL resources.

The TFC algorithm models deadlocks using the Channel Dependency Graph (CDG) and partitions \name's  topology into subgraphs. Within each subgraph, topology steering rules and VL constraints are unified into a single set. It applies an N-dimensional cross-dimensional loop-breaking principle to decompose the set into single-VL acyclic subsets, followed by a same-dimensional loop-breaking principle to compute permutations and Cartesian products of power set elements. This generates all-path combinations and VL mappings, ensuring deadlock-free operation. Due to page limits, detailed explanations are omitted.

\begin{figure} [t]
    \centering
    \includegraphics[width=0.98\linewidth]{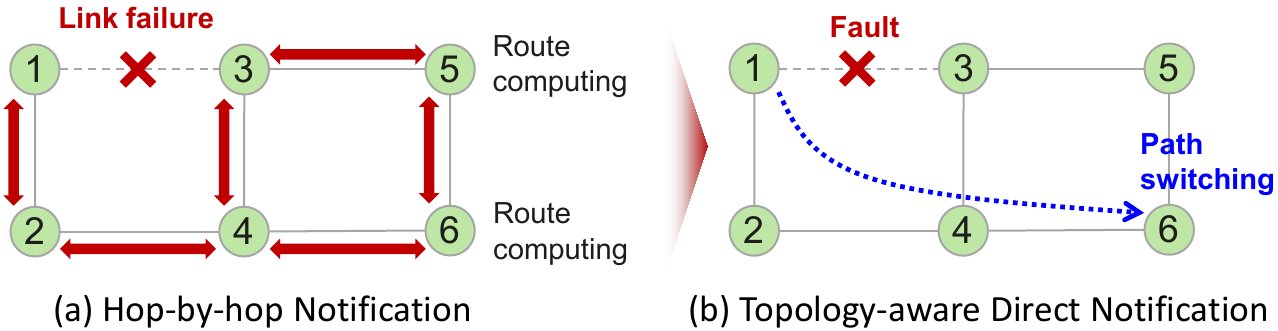} 
    \caption{{From Hop-by-hop to Direct Notification}}
         \label{fig:fault_tolerance}
\end{figure}

\subsection{Fast Fault Recovery via Direct Notification} 
When link failures occur in a network, traditional routing system usually adopt  hop-by-hop notification. As illustrated by the left figure in Fig.~\ref{fig:fault_tolerance}, when link 1-3 has a failure, such an information is propagated by node 1 and 3 hop-by-hop, which usually consumes a long latency. 
In \name,  since each node has a deterministic set of communication target, we can accelerate the routing convergence by directly notifying those nodes upon link failures.  As shown by the right figure, when link 1-3 has failure, such information is directly sent to node 6 according to pre-computed routing relationships.  With such a topology-aware direct notification, the control plane overhead can be greatly reduced.

\section{Maximizing The Performance}
\label{sec:system_perf_opt}
\begin{figure} [t]
    \centering
    \includegraphics[width=0.99\linewidth]{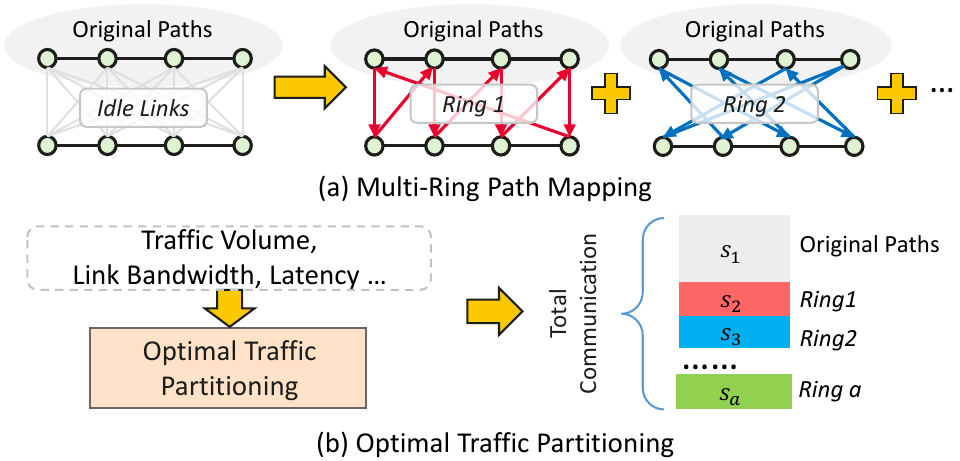} 
    \caption{{ Multi-Ring Algorithm for \texttt{AllReduce}}}
         \label{fig:multiring}
\end{figure}

\begin{figure} [t]
    \centering
    \includegraphics[width=0.99\linewidth]{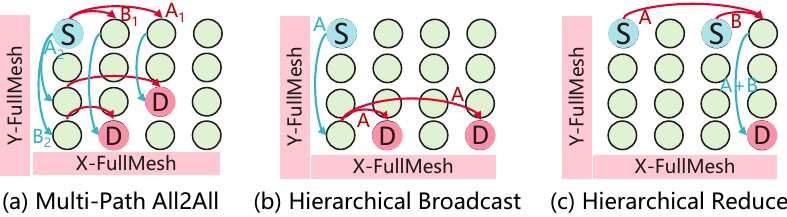} 
    \caption{{Multi-Path and Hierarchical All-to-All}}
         \label{fig:all2all}
\end{figure}

Although \name's hardware architecture design matches the traffic patterns of LLM training, running workloads on such a hierarchical AI cluster may suffer from low utilization if training tasks are not effectively distributed across computing resources. Additionally, the performance of collective communication is critical for overall training efficiency. To ensure optimal system performance during LLM training, we introduce several topology-aware optimization strategies to  enhance system performance further.

\subsection{Topology-Aware Collective Communication}

\label{sec:cc_algorithm}

To optimize collective communication on \name~and fully utilize the bandwidth of hierarchical direct interconnects, we present topology-aware collective communication algorithms leveraging the All-Path-Routing capability of \name~ (Sec.~\ref{sec:apr}). We take \texttt{All-Reduce} and \texttt{All-to-All} examples to convey our concepts:

\textbf{\texttt{All-Reduce}}:  We propose a \emph{Multi-Ring} algorithm for efficient \texttt{AllReduce} on \name. We begin by modeling the network topology in a unified abstraction, accounting for factors such as node count, inter-node connections, link bandwidth, and latency. Next, we integrate collective communication with path mapping using a logical multi-ring algorithm, ensuring exclusive path usage without traffic conflicts. As depicted in Fig.~\ref{fig:multiring}-(a), the \emph{Original Paths} represent default mappings. Idle links, excluded from these paths, are leveraged via the APR mechanism to enhance bandwidth. Finally, as shown in Fig.~\ref{fig:multiring}-(b), we optimize traffic partitioning across multiple paths to mitigate bottlenecks and maximize the benefits of borrowed bandwidth through APR.

\textbf{\texttt{All-to-All}}:
In the scenario of \texttt{All-to-All} communication, we consider two representative use cases: 

(1) General \texttt{All-to-All:} As shown in Fig.~\ref{fig:all2all}-(a), when the source node in \name~(we demonstrate a 2D mesh in the figure for simplicity. The proposed techniques can be extended to higher-dimensions) sends different data to different  destination nodes, we adopt a \emph{Multi-Path All2All} optimization. Specifically, each element (vector or tensor) is split into two partitions, which are transmitted simultaneously along the X-FullMesh and Y-FullMesh interconnects and use at most one-hop forwarding to arrive at the destination. Such a strategy guarantees a high bandwidth utilization in \name's nD-FullMesh architecture.

(2) Broadcast + Reduce: For \texttt{All-to-All} operations involving token distribution and expert data collection~\cite{deepseekai2024deepseekv3technicalreport}, the semantics are equivalent to overlapping multiple \texttt{broadcast} and \texttt{reduce} operations. As shown in Fig.~\ref{fig:all2all}-(b) and (c), we can adopt hierarchical broadcast/reduce to save bandwidth usage,  fully leveraging \name's hierarchical topology.

\subsection{Topology-Aware Parallelization}
\label{sec:cc}

\begin{figure} [t]
    \centering
    \includegraphics[width=0.99\linewidth]{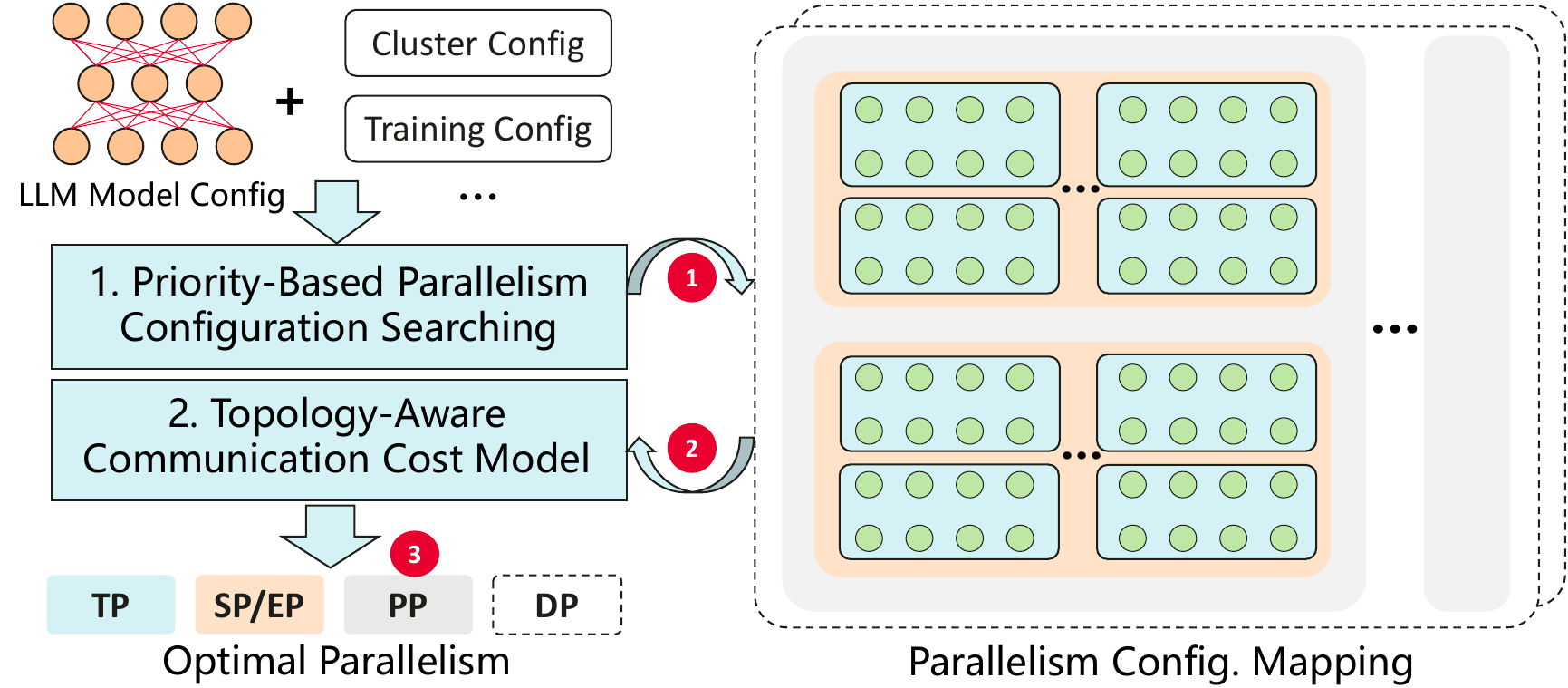} 
    \caption{{Topology-Aware Parallelization}}
         \label{fig:topology_aware_parallel}
\end{figure}

Given a training task and system, we determine optimal parallelization strategies to fully utilize the high-bandwidth local interconnects of \name. We adopt a Topology-Aware Parallelization mechanism to optimize model and data splitting for LLM training.
As shown in Fig.~\ref{fig:topology_aware_parallel}, our approach involves:

\noindent\emph{Step} \textcircled{1}: Generating feasible parallelism configurations and mapping them onto the UB-Mesh architecture.

\noindent\emph{Step} \textcircled{2}: Evaluating communication costs with the Topology-Aware Communication Cost Model.

\noindent\emph{Step} \textcircled{3}: Iteratively minimizing communication overhead to find the optimal configuration.

This mechanism has two important requirements: (1) Constructing a proper search space to balance efficiency and performance. (2) Ensuring the cost model is as accurate as possible.
For requirement (1), we prune the search space using a priority-based heuristic: TP and SP (or CP), which involve high communication volumes, are prioritized for high-bandwidth domains, while PP and DP, used during gradient updates, is the lowest priority. For MoE models requiring EP, we force SP*DP as an integer multiple of EP.
For requirement (2), we accurately model the behavior of APR and Topology-Aware Collective Communication on \name's topology and use an accurate in-house simulation infrastructure to calibrate the model.

\section{Evaluation}
\label{sec:eval}

In this section, we explore the architecture design space of \name~and analyze its advantages against baseline Clos architecture. To facilitate a more detailed comparison, we divide our analysis into two tiers: intra-rack architecture comparison and inter-rack architecture comparison.

\subsection{Experiments Setup}

\begin{table}[t]
    \caption{Benchmarking Models}
    \label{tab:benchmark}
    \resizebox{0.48\textwidth}{!}{\renewcommand{\arraystretch}{1.0}{
        \begin{tabular}{|c|c|c|c|c|c|c|}
            \hline
                                                                                     & Name        & \#Layers & \#Heads & Head Size & Hidden Dim & \#Experts \\ \hline
            \multirow{3}{*}{\begin{tabular}[c]{@{}c@{}}Dense\\ Models\end{tabular}}  & LLAMA-70B   & 80       & 64      & 128       & 8192       & /         \\ \cline{2-7} 
                                                                                     & GPT3-175B   & 96       & 96      & 128       & 12288      & /         \\ \cline{2-7} 
                                                                                     & Dense-1T    & 128      & 128     & 192       & 24576      & /         \\ \hline
            \multirow{2}{*}{\begin{tabular}[c]{@{}c@{}}MoE\\ Models\end{tabular}} & GPT4-2T *   & 96       & 96      & 128       & 12288      & 16        \\ \cline{2-7} 
                                                                                     & MoE-10T  & 128      & 144 & 128   & 18432      & 32        \\ \hline
            \end{tabular}
    }}

\end{table}

 Table~\ref{tab:benchmark} lists the benchmarking workloads: LLAMA-70B and GPT3-175B are dense models while GPT4-2T is a sparse  model adopting MoE techniques~\cite{ds-moe}. Note that GPT4's architecture has not been released officially, we adopt the inferred parameters~\cite{semianalysis2023gpt4}. To evaluate system's performance on larger models, we also include Dense-1T and MoE-10T models.
To explore the architecture of \name, we build an in-house simulation infrastructure, which has been aligned with the real PoC hardware,  to evaluate the cluster-scale LLM training performance. 

\subsection{Intra-Rack Architecture Exploration}

We compare among different intra-rack network architectures, which are illustrated in Fig.~\ref{fig:baseline_intra_rack}:

\begin{figure} [t]
    \centering
    \includegraphics[width=0.98\linewidth]{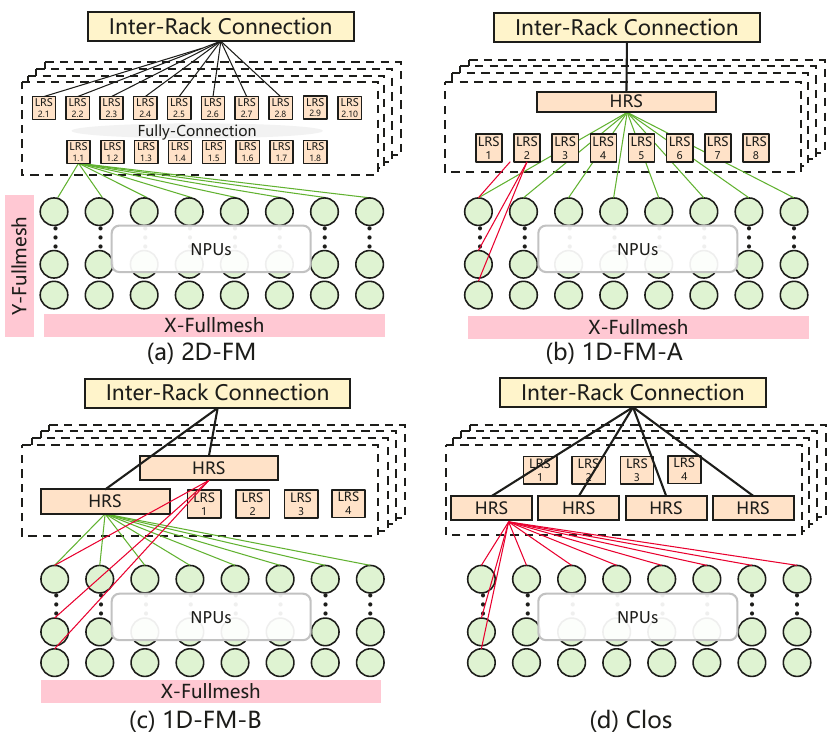} 
    \caption{{Baseline Intra-Rack Architectures}}
         \label{fig:baseline_intra_rack}
\end{figure}

\textbf{(a) 2D-FM:}  \name's architecture. 64 NPUs are directly interconnected via electrical cables, forming a 2D full-mesh network topology (X-Fullmesh + Y-Fullmesh). This design offers low cost and adopts  72 LRS (see Table.~\ref{tab:building_blocks}) for aggregating inter-rack bandwidth and CPU-NPU interconnection (CPUs are omitted in the figure for simplicity).  

\textbf{(b) 1D-FM-A:} This alternative architecture retains the 1D X-Fullmesh, namely the 8 NPUs on each board are still directly interconnected. However, cross-board communication is realized via 32 LRS. Each NPU has a UB x16 IO connecting to LRS. Another UB x16 IO are connected to four high-radix switches (HRS) for inter-rack communication.  

\textbf{(c) 1D-FM-B:} This architecture replaces LRS with HRS further. Four LRS in each backplane is used for NPU-to-CPU communication. 
Cross-board NPU communication is realized via eight switches in four backplanes. These switches are also connected to inter-rack networks, offering a UB x32 IO for inter-rack communication for each NPU.

\textbf{(d) Clos:}  This architecture does not use inter-NPU direct connections, instead, it connects all ports of 64 NPUs  to $4\times4$ HRS, forming a symmetrical Clos topology. This architecture provides the highest flexibility but also requires lots of switching resources. 

\begin{figure} [t]
    \centering
    \includegraphics[width=0.96\linewidth]{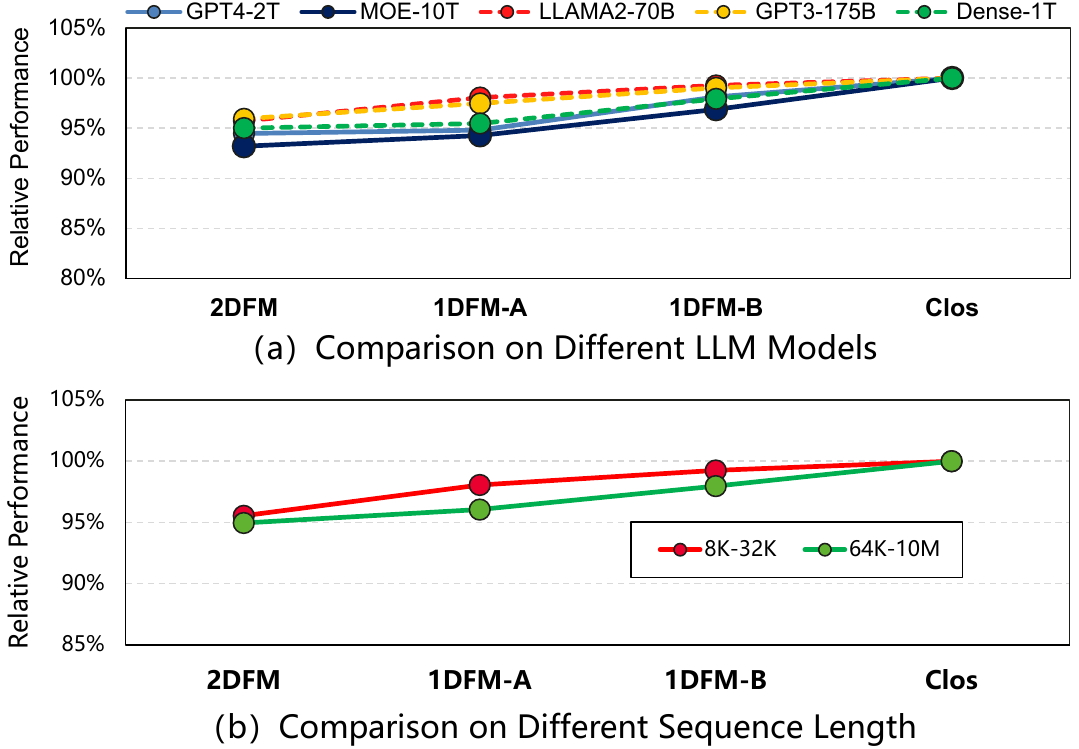} 
    \caption{{Performance of Different Intra-Rack Topology}}
         \label{fig:intra-rack-topology}
\end{figure}

\textbf{Performance Comparison:} In  Fig.~\ref{fig:intra-rack-topology}, we fix the inter-rack architecture (2D-FM, see Sec.\ref{sec:inter-rack-arch})  and compare the training throughput of various intra-rack architectures, relative to the Clos baseline. 
The SuperPod scale is 8K (128 racks), and the sequence lengths evaluated range from 8K to 10M. We calculate the average performance among different sequence lengths.

As shown in Fig.~\ref{fig:intra-rack-topology}-(a), compared to the  Clos architecture, the 2D-FM architecture achieves 93.2\% to 95.9\% of the training performance.
The 1D-FM-A architecture shows lower performance degradation, achieving a performance improvement of 2.44\% for LLAMA2-70B compared to 2D-FM. For other models with more parameters, the improvement is less than 1.6\%. The 1D-FM-B architecture demonstrates a slightly higher performance improvement of more than 3\% compared to the 2D-FM architecture thanks to higher inter-rack bandwidth, but the improvement is still marginal.

Fig.~\ref{fig:intra-rack-topology}-(b) explores the performance on different sequence lengths, relative to the baseline scenario of the Clos architecture (all-models average). For sequence lengths ranging from 8K to 32K, the 2D-FM architecture achieves  95.5\% of the performance, which is slightly lower than 1DFM-A (98.1\%) and 1DFM-B (99.2\%).  For sequence lengths from 64K to 10M, the 2D-FM architecture achieves 95.0\% of the performance compared to the Clos architecture.

 We can see that {\textbf{compared to Clos, 2D-FM architecture  offers similar training performance (performance gap within 7\%)} with much lower  hardware cost}, which will be evaluated in Sec.~\ref{sec:cost-eff}.

\subsection{Inter-Rack Architecture Exploration}
\label{sec:inter-rack-arch}
Beyond the intra-rack 2D-FM, \name~adopts another 2D-FM within the \name~Pod, forming a 4D-FM network architecture. 
As illustrated Fig.~\ref{fig:inter-rack-arch-compare}, we compare \name's 2D full-mesh inter-rack architecture with the baseline Clos architecture:
\begin{figure} [t]
    \centering
    \includegraphics[width=0.99\linewidth]{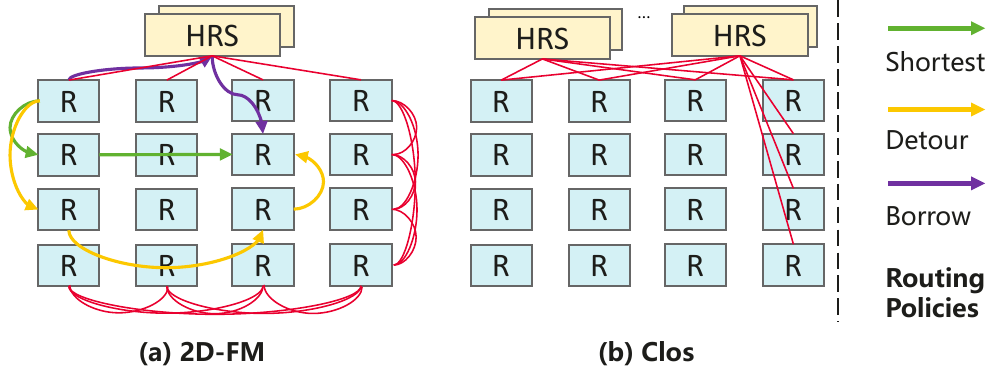} 
    \caption{{Inter-Rack Network Architectures}}
         \label{fig:inter-rack-arch-compare}
\end{figure}

\textbf{(a) 2D-FM:} The 16 racks are directly interconnected horizontally and vertically, forming a 2D full-mesh direct connection. All racks are connected to HRS switches, enabling cross-pod interconnection.   

In \name-Pod, three routing strategies can be supported: 

\begin{itemize}[leftmargin=1em]

\item\emph{ Shortest: } The baseline routing strategy, only the shortest path on the 2D mesh will be selected for P2P communication. 

\item  \emph{ Detour: } According to the All-Path-Routing mechanism, other paths will also be used for maximizing the bandwidth. 

\item \emph{ Borrow: }  Since all racks are also connected to switches, we enable racks to "borrow" the bandwidth from switches. 
\end{itemize}

\textbf{(b) Clos:} All racks are connected to HRS switches, without direct connection. This architecture consumes many more switches than 2D-FM but provides the highest all-to-all bandwidth and flexibility.

\begin{figure} [t]
    \centering
    \includegraphics[width=0.98\linewidth]{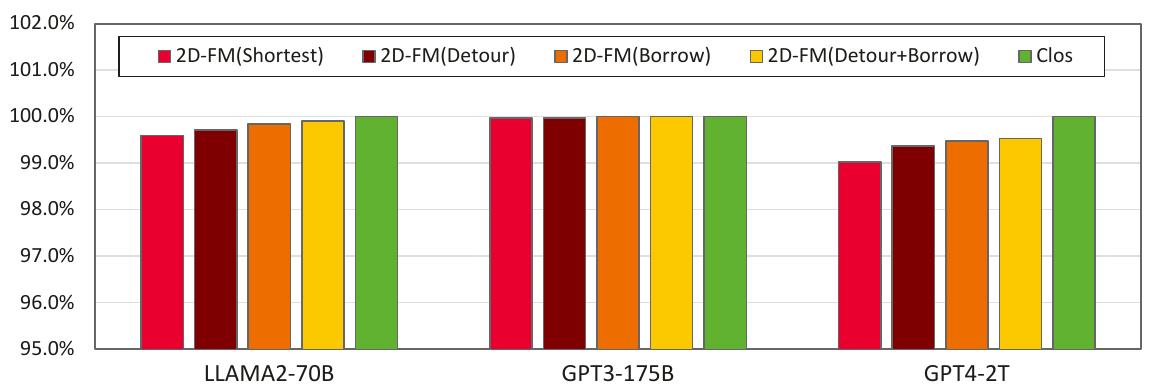} 
    \caption{{End-to-End Performance Comparison of Different Inter-Rack Interconnections}}
         \label{fig:inter-rack-e2e}
\end{figure}

\textbf{Performance Comparison:}  As shown in Fig.~\ref{fig:inter-rack-e2e},  we compare the performance of 2D-FM (including different routing strategies) and Clos architectures. As we can see, compared to the ideal Clos architecture,  the performance gap between 2D-FM and Clos is negligible, especially when Detour and Borrow strategies are applied. The GPT3-175B is not sensitive to inter-rack communication performance, while on GPT4-2T, the naive 2D-FM with shortest-path routing suffers from 0.73\% training performance degradation. The performance gap is narrowed to merely 0.46\% when \emph{Detour} and \emph{Borrow} routing strategies are adopted.  Generally, \textbf{the 2D-FM inter-rack interconnects demonstrates almost the same performance as the expensive Clos architecture. }

\begin{figure} [t]
    \centering
    \includegraphics[width=0.98\linewidth]{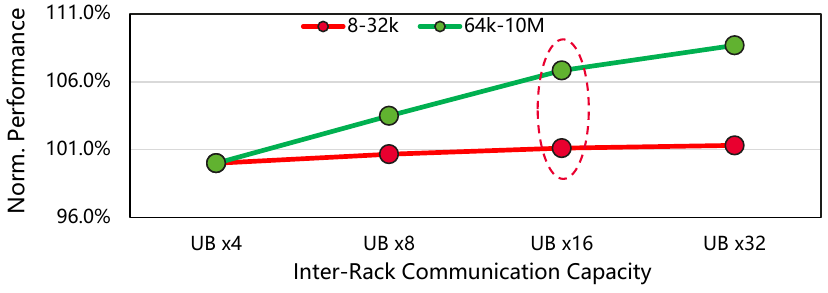} 
    \caption{{Inter-Rack   Bandwidth Exploration}}
         \label{fig:exploration}
\end{figure}

{\textbf{Inter-Rack Bandwidth Exploration:}} Fig.~\ref{fig:exploration}-(b)  compares the throughput performance of 8K SuperPod under varying inter-rack bandwidth conditions.  The inter-rack bandwidths evaluated are x4, x8, x16, and x32 UB IO per NPU. For sequence lengths ranging from 8K to 32K, the optimal inter-rack bandwidth is identified as UB x16, while for sequence lengths from 64K to 10M, the optimal inter-rack bandwidth is identified as UB x32. The performance gain from increasing the inter-rack bandwidth from UB x8 to UB x16 in the 8K-32K sequence length range is minimal, at only 0.44\%. However, the performance gain from increasing the inter-rack bandwidth from UB x16 to UB x32 in the 64K-10M sequence length range is more significant, at 1.85\%.  In scenarios with sequence lengths from 64K to 10M, a portion of the TP (Tensor Parallelism) and SP (Sequence Parallelism) traffic inevitably traverses the inter-rack link. Higher inter-rack bandwidths significantly reduce the communication time for TP and SP, leading to a more pronounced performance improvement in these scenarios. The data underscores the importance of matching inter-rack bandwidth to the specific sequence length requirements of different model scenarios, particularly in reducing TP and SP communication latency for large-scale models. UB-Mesh  assigns a UB x16 IO per NPU for inter-rack communication by default to achieve a balance between cost and performance. \textbf{We can also adjust the intra/inter-rack bandwidth ratio to match the specific requirements from certain LLM training workloads.}

\subsection{Cost-Efficiency Comparison} 
\label{sec:cost-eff}

The system cost is usually measured in TCO, i.e., \textit{TCO = CapEx + OpEx}. Considering the achieved training performance relative to the baseline, we define the cost-efficiency of a system as following:
\begin{equation}
    \label{eq:cost_eff}
    \textit{Cost-efficiency} = \frac{\textit{Average Performance}}{\texttt{OpEx + CapEx}}
\end{equation}

\begin{figure} [t]
    \centering
    \includegraphics[width=0.98\linewidth]{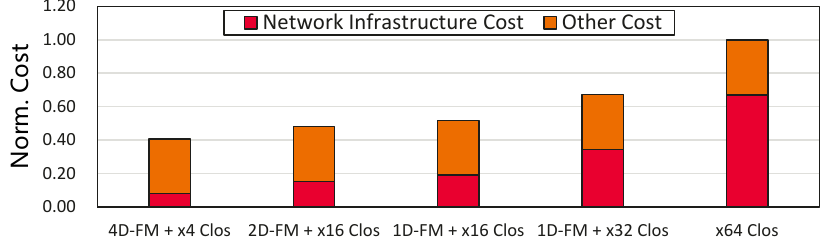} 
    \caption{{CapEx  Comparison}}
         \label{fig:cost}
\end{figure}

We estimate the CapEx cost, including the costs of NPUs, CPUs, LRS, HRS, cables and other modules using in-house numbers, and compare among different architectures. 
 As shown in Fig.~\ref{fig:cost}, \name's 4D-FM+Clos architecture achieves $1.18\times$, $1.26\times$,  $1.65\times$ and $2.46\times$ CapEx reduction compared to 2D-FM+x16 (denotes UB x16 IO per-NPU) Clos, 1D-FM+x16 Clos, and x64T Clos architectures, respectively. Compared to the baseline Clos architecture, \name~successfully reduces the ratio of network infrastructure cost in the system from 67\% to 20\%, due to the saving of high-performance switches and long-range optical cables/modules. According to our evaluation, 98\% of high-radix switches and 93\% of optical modules are saved compared to baseline Clos architecture.  
 
The OpEx reduction is mainly composed of electricity bill and maintenance cost in the lifetime of the system. UB-mesh reduces OpEx by about 35\% compared with Clos, due to its much fewer use of switches and optic modules. According to the estimation of AI system by our Cloud Division, OpEx accounts for around 30\% of TCO. Ultimately, \name~achieves $\costsaving\times$ higher cost-efficiency according to equation~\ref{eq:cost_eff}. 

\subsection{Linearity Evaluation}

The linearity of an AI cluster refers to the extent to which the performance of the cluster scales linearly with the number of NPUs. 
Specifically, 
we measure the linearity  by the following equation:
\begin{equation}
    \textit{Linearity} = \frac{\textit{Per-NPU Perf. on Target Scale}}{ \textit{Per-NPU Perf. on Base Scale}}\times 100\%
\end{equation}

\begin{figure} [t]
    \centering
    \includegraphics[width=0.99\linewidth]{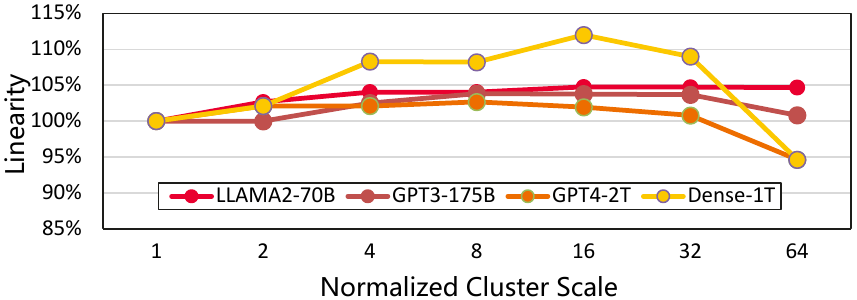} 
    \caption{{Linearity Analysis @ Sequence 256K}}
         \label{fig:linearity}
\end{figure}

Fig.~\ref{fig:linearity} evaluates the linearity of \name~under varying cluster scales. 
The base scales (namely, the $1\times$ scale in the figure) also vary among tasks. Specifically, the  LLAMA2-70B uses 128 NPUs, GPT3-175B's base scale is 512, Dense-1T and GPT4-2T use 1K NPUs.

 As we can see, the linearity of \name~on all tasks exceeds 100\% under 1$\times$ to 32$\times$ scales, this is because improving the scale provides more high-bandwidth domains and unlocks the potential of searching for a better parallelism strategy to improve MFU. The linearity of GPT4-2T and Dense-1T models drop when under 64$\times$ scale with 64K NPUs involved, but the linearity is still above 95\%.

\subsection{Network Reliability Analysis}  
AI training procedures are sensitive to hardware failures. A classic time-based metric is used to evaluate the reliability of a system: 
 \begin{equation}
    \label{eq:availability}
    Availability = \frac{MTBF}{MTBF+MTTR} \times 100\%
 \end{equation}
Where MTBF denotes Mean Time Between Failure and MTTR refers
to Meam Time to Repair. 

As estimated in Table~\ref{tab:health}, prioritizing the usage of direct-connected electrical cables (E-Cables) over optical fibers and switches result in greatly reduced Annualized Failure Rate (AFR) of network modules. 
 Then we can calculate the MTBF ($MTBF=\frac{365\times24}{AFR}$) of two architectures: the baseline Clos is 13.8 hours for a 8K-NPU cluster, while UB-Mesh achieves 98.5 hours, a 7.14$\times$ improvement. 
Ultimately, the availability of UB-Mesh is 98.8\% according to Eq.\ref{eq:availability} ({We assume a 75-minute MTTR according to our existing statistics}), significantly outperforming Clos's 91.6\% (7.2\% improvement). 

To further improve availability, we also carefully craft an in-house network monitoring tools, which can quickly identify and locate network failures within 10 minutes and then trigger timely task migration within 3 minutes, greatly reducing MTTR. With such an MTTR optimization, the availability of \name~can be further improved to 99.78\% according our evaluation.

 \begin{table}[t]
    \caption{MTBF Estimation}
    \label{tab:health}
    \resizebox{0.48\textwidth}{!}{\renewcommand{\arraystretch}{1.3}{
        \begin{tabular}{|l|ccccc|c|}
            \hline
            \multirow{2}{*}{}             & \multicolumn{5}{c|}{Annualized Failure Rate (AFR)}                                                                                                 & \multirow{2}{*}{\begin{tabular}[c]{@{}c@{}}MTBF\\ (Hours)\end{tabular}} \\ \cline{2-6}
                                          & \multicolumn{1}{c|}{Electrical Cable} & \multicolumn{1}{c|}{Optical Cables} & \multicolumn{1}{c|}{LRS} & \multicolumn{1}{c|}{HRS}  & Total Failure &                                                                         \\ \hline
            \multicolumn{1}{|c|}{UB-Mesh} & \multicolumn{1}{c|}{5.82}             & \multicolumn{1}{c|}{1.55}           & \multicolumn{1}{c|}{81}  & \multicolumn{1}{c|}{0.56} & 88.9          & 98.5                                                                    \\ \hline
            \multicolumn{1}{|c|}{Clos}    & \multicolumn{1}{c|}{13.8}             & \multicolumn{1}{c|}{574}            & \multicolumn{1}{c|}{18}  & \multicolumn{1}{c|}{27}   & 632.8         & 13.8                                                                    \\ \hline
            \end{tabular}
    }}
    \end{table}

  \textbf{Summary:} Compared to the baseline  Clos architectures, \name~achieves \textbf{\costsaving$\times$ higher cost-efficiency} at the cost of marginal performance degradation (within 7\%). \name~\textbf{improves network availability by 7.2\%} due to the greatly reduced usage of switches and optical modules.  \name~also achieves \textbf{95\%+ linearity} on several LLM training tasks.  
\section{Discussion}

\noindent\textbf{Co-Processor for Collective Communication.} It is worth mentioning that, 
in the UB IO controller there is also a special co-processor called Collective Communication Unit (CCU) for offloading the collective communication tasks. Specifically, CCU executes instructions, actively reads/writes data from/to HBM, initiates inter-NPU transmission and also performs in-line data reduce utilizing on-chip SRAM buffer. It eliminates the redundant data copy from application's memory buffer to the communication buffer.  
 Such a design effectively alleviates the HBM bandwidth consumption  and maintains the deterministic reduce order via a checkbit-based fine-grained synchronization mechanism.  CCU can also seamlessly co-operate with compute cores to achieve efficient compute-communication overlap.

\noindent\textbf{Support for Massive-Expert Models.}
Besides dense and regular MoE models, we notice that people are actively exploring MoE models containing massive experts~\cite{deepseekai2024deepseekv3technicalreport, deepseekai2024deepseekv2strongeconomicalefficient}. Such a design usually calls for  large-scale, fine-grained \texttt{all-to-all} communication. \name~is able to efficiently support such massive-expert models via  the multi-path and hierarchical all-to-all optimization~(Sec. \ref{sec:cc_algorithm}) and UB's Load/Store based data transfer. The CCU can also efficiently offload  \texttt{all-to-all} operations to save the usage of precious compute cores, as expected in \cite{deepseekai2024deepseekv3technicalreport}.
\section{Conclusion}

This paper proposes \name, a novel datacenter network architecture designed for next-generation LLM training.  \name~adopts an nD-FullMesh network topology to save the usage of switches and optical models, while adapts to the hierarchically localized traffic patterns of LLM workloads. We carefully consider the architecture, physical implementation and networking system optimization and propose several techniques to overcome various challenges. Compared to traditional Clos networks, \name~achieves similar LLM training performance while provides \costsaving$\times$ higher system-level cost efficiency and 7.2\% higher availability. \name~also achieves 95\%+ linearity on several LLM training tasks.


\bibliographystyle{ACM-Reference-Format}
\bibliography{refs}


\begin{thebibliography}{32}


\ifx \showCODEN    \undefined \def \showCODEN     #1{\unskip}     \fi
\ifx \showDOI      \undefined \def \showDOI       #1{#1}\fi
\ifx \showISBNx    \undefined \def \showISBNx     #1{\unskip}     \fi
\ifx \showISBNxiii \undefined \def \showISBNxiii  #1{\unskip}     \fi
\ifx \showISSN     \undefined \def \showISSN      #1{\unskip}     \fi
\ifx \showLCCN     \undefined \def \showLCCN      #1{\unskip}     \fi
\ifx \shownote     \undefined \def \shownote      #1{#1}          \fi
\ifx \showarticletitle \undefined \def \showarticletitle #1{#1}   \fi
\ifx \showURL      \undefined \def \showURL       {\relax}        \fi
\providecommand\bibfield[2]{#2}
\providecommand\bibinfo[2]{#2}
\providecommand\natexlab[1]{#1}
\providecommand\showeprint[2][]{arXiv:#2}

\bibitem[AI et~al\mbox{.}(2024)]%
        {ai2024yiopenfoundationmodels}
\bibfield{author}{\bibinfo{person}{01. AI}, \bibinfo{person}{:}, \bibinfo{person}{Alex Young}, \bibinfo{person}{Bei Chen}, \bibinfo{person}{Chao Li}, \bibinfo{person}{Chengen Huang}, \bibinfo{person}{Ge Zhang}, \bibinfo{person}{Guanwei Zhang}, \bibinfo{person}{Heng Li}, \bibinfo{person}{Jiangcheng Zhu}, \bibinfo{person}{Jianqun Chen}, \bibinfo{person}{Jing Chang}, \bibinfo{person}{Kaidong Yu}, \bibinfo{person}{Peng Liu}, \bibinfo{person}{Qiang Liu}, \bibinfo{person}{Shawn Yue}, \bibinfo{person}{Senbin Yang}, \bibinfo{person}{Shiming Yang}, \bibinfo{person}{Tao Yu}, \bibinfo{person}{Wen Xie}, \bibinfo{person}{Wenhao Huang}, \bibinfo{person}{Xiaohui Hu}, \bibinfo{person}{Xiaoyi Ren}, \bibinfo{person}{Xinyao Niu}, \bibinfo{person}{Pengcheng Nie}, \bibinfo{person}{Yuchi Xu}, \bibinfo{person}{Yudong Liu}, \bibinfo{person}{Yue Wang}, \bibinfo{person}{Yuxuan Cai}, \bibinfo{person}{Zhenyu Gu}, \bibinfo{person}{Zhiyuan Liu}, {and} \bibinfo{person}{Zonghong Dai}.} \bibinfo{year}{2024}\natexlab{}.
\newblock \bibinfo{title}{Yi: Open Foundation Models by 01.AI}.
\newblock
\newblock
\showeprint[arxiv]{2403.04652}~[cs.CL]
\urldef\tempurl%
\url{https://arxiv.org/abs/2403.04652}
\showURL{%
\tempurl}


\bibitem[Ajima et~al\mbox{.}(2009)]%
        {ajima2009tofu}
\bibfield{author}{\bibinfo{person}{Yuichiro Ajima}, \bibinfo{person}{Shinji Sumimoto}, {and} \bibinfo{person}{Toshiyuki Shimizu}.} \bibinfo{year}{2009}\natexlab{}.
\newblock \showarticletitle{Tofu: A 6D mesh/torus interconnect for exascale computers}.
\newblock \bibinfo{journal}{\emph{Computer}} \bibinfo{volume}{42}, \bibinfo{number}{11} (\bibinfo{year}{2009}), \bibinfo{pages}{36--40}.
\newblock


\bibitem[An et~al\mbox{.}(2024)]%
        {firefly}
\bibfield{author}{\bibinfo{person}{Wei An}, \bibinfo{person}{Xiao Bi}, \bibinfo{person}{Guanting Chen}, \bibinfo{person}{Shanhuang Chen}, \bibinfo{person}{Chengqi Deng}, \bibinfo{person}{Honghui Ding}, \bibinfo{person}{Kai Dong}, \bibinfo{person}{Qiushi Du}, \bibinfo{person}{Wenjun Gao}, \bibinfo{person}{Kang Guan}, \bibinfo{person}{Jianzhong Guo}, \bibinfo{person}{Yongqiang Guo}, \bibinfo{person}{Zhe Fu}, \bibinfo{person}{Ying He}, \bibinfo{person}{Panpan Huang}, \bibinfo{person}{Jiashi Li}, \bibinfo{person}{Wenfeng Liang}, \bibinfo{person}{Xiaodong Liu}, \bibinfo{person}{Xin Liu}, \bibinfo{person}{Yiyuan Liu}, \bibinfo{person}{Yuxuan Liu}, \bibinfo{person}{Shanghao Lu}, \bibinfo{person}{Xuan Lu}, \bibinfo{person}{Xiaotao Nie}, \bibinfo{person}{Tian Pei}, \bibinfo{person}{Junjie Qiu}, \bibinfo{person}{Hui Qu}, \bibinfo{person}{Zehui Ren}, \bibinfo{person}{Zhangli Sha}, \bibinfo{person}{Xuecheng Su}, \bibinfo{person}{Xiaowen Sun}, \bibinfo{person}{Yixuan Tan}, \bibinfo{person}{Minghui Tang}, \bibinfo{person}{Shiyu Wang}, \bibinfo{person}{Yaohui Wang}, \bibinfo{person}{Yongji Wang}, \bibinfo{person}{Ziwei Xie}, \bibinfo{person}{Yiliang Xiong}, \bibinfo{person}{Yanhong Xu}, \bibinfo{person}{Shengfeng Ye}, \bibinfo{person}{Shuiping Yu}, \bibinfo{person}{Yukun Zha}, \bibinfo{person}{Liyue Zhang}, \bibinfo{person}{Haowei Zhang}, \bibinfo{person}{Mingchuan Zhang}, \bibinfo{person}{Wentao Zhang}, \bibinfo{person}{Yichao Zhang}, \bibinfo{person}{Chenggang Zhao}, \bibinfo{person}{Yao Zhao}, \bibinfo{person}{Shangyan Zhou}, \bibinfo{person}{Shunfeng Zhou}, {and} \bibinfo{person}{Yuheng Zou}.} \bibinfo{year}{2024}\natexlab{}.
\newblock \bibinfo{title}{Fire-Flyer AI-HPC: A Cost-Effective Software-Hardware Co-Design for Deep Learning}.
\newblock
\newblock
\showeprint[arxiv]{2408.14158}~[cs.DC]
\urldef\tempurl%
\url{https://arxiv.org/abs/2408.14158}
\showURL{%
\tempurl}


\bibitem[Bai et~al\mbox{.}(2023)]%
        {bai2023qwentechnicalreport}
\bibfield{author}{\bibinfo{person}{Jinze Bai}, \bibinfo{person}{Shuai Bai}, \bibinfo{person}{Yunfei Chu}, \bibinfo{person}{Zeyu Cui}, \bibinfo{person}{Kai Dang}, \bibinfo{person}{Xiaodong Deng}, \bibinfo{person}{Yang Fan}, \bibinfo{person}{Wenbin Ge}, \bibinfo{person}{Yu Han}, \bibinfo{person}{Fei Huang}, \bibinfo{person}{Binyuan Hui}, \bibinfo{person}{Luo Ji}, \bibinfo{person}{Mei Li}, \bibinfo{person}{Junyang Lin}, \bibinfo{person}{Runji Lin}, \bibinfo{person}{Dayiheng Liu}, \bibinfo{person}{Gao Liu}, \bibinfo{person}{Chengqiang Lu}, \bibinfo{person}{Keming Lu}, \bibinfo{person}{Jianxin Ma}, \bibinfo{person}{Rui Men}, \bibinfo{person}{Xingzhang Ren}, \bibinfo{person}{Xuancheng Ren}, \bibinfo{person}{Chuanqi Tan}, \bibinfo{person}{Sinan Tan}, \bibinfo{person}{Jianhong Tu}, \bibinfo{person}{Peng Wang}, \bibinfo{person}{Shijie Wang}, \bibinfo{person}{Wei Wang}, \bibinfo{person}{Shengguang Wu}, \bibinfo{person}{Benfeng Xu}, \bibinfo{person}{Jin Xu}, \bibinfo{person}{An Yang}, \bibinfo{person}{Hao Yang}, \bibinfo{person}{Jian Yang}, \bibinfo{person}{Shusheng Yang}, \bibinfo{person}{Yang Yao}, \bibinfo{person}{Bowen Yu}, \bibinfo{person}{Hongyi Yuan}, \bibinfo{person}{Zheng Yuan}, \bibinfo{person}{Jianwei Zhang}, \bibinfo{person}{Xingxuan Zhang}, \bibinfo{person}{Yichang Zhang}, \bibinfo{person}{Zhenru Zhang}, \bibinfo{person}{Chang Zhou}, \bibinfo{person}{Jingren Zhou}, \bibinfo{person}{Xiaohuan Zhou}, {and} \bibinfo{person}{Tianhang Zhu}.} \bibinfo{year}{2023}\natexlab{}.
\newblock \bibinfo{title}{Qwen Technical Report}.
\newblock
\newblock
\showeprint[arxiv]{2309.16609}~[cs.CL]
\urldef\tempurl%
\url{https://arxiv.org/abs/2309.16609}
\showURL{%
\tempurl}


\bibitem[Clark et~al\mbox{.}(2022)]%
        {clark2022unified}
\bibfield{author}{\bibinfo{person}{Aidan Clark}, \bibinfo{person}{Diego de Las~Casas}, \bibinfo{person}{Aurelia Guy}, \bibinfo{person}{Arthur Mensch}, \bibinfo{person}{Michela Paganini}, \bibinfo{person}{Jordan Hoffmann}, \bibinfo{person}{Bogdan Damoc}, \bibinfo{person}{Blake Hechtman}, \bibinfo{person}{Trevor Cai}, \bibinfo{person}{Sebastian Borgeaud}, {et~al\mbox{.}}} \bibinfo{year}{2022}\natexlab{}.
\newblock \showarticletitle{Unified scaling laws for routed language models}. In \bibinfo{booktitle}{\emph{International conference on machine learning}}. PMLR, \bibinfo{pages}{4057--4086}.
\newblock


\bibitem[DeepSeek-AI et~al\mbox{.}(2024a)]%
        {deepseekai2024deepseekv2strongeconomicalefficient}
\bibfield{author}{\bibinfo{person}{DeepSeek-AI}, \bibinfo{person}{Aixin Liu}, \bibinfo{person}{Bei Feng}, \bibinfo{person}{Bin Wang}, \bibinfo{person}{Bingxuan Wang}, \bibinfo{person}{Bo Liu}, \bibinfo{person}{Chenggang Zhao}, \bibinfo{person}{Chengqi Dengr}, \bibinfo{person}{Chong Ruan}, \bibinfo{person}{Damai Dai}, \bibinfo{person}{Daya Guo}, \bibinfo{person}{Dejian Yang}, \bibinfo{person}{Deli Chen}, \bibinfo{person}{Dongjie Ji}, \bibinfo{person}{Erhang Li}, \bibinfo{person}{Fangyun Lin}, \bibinfo{person}{Fuli Luo}, \bibinfo{person}{Guangbo Hao}, \bibinfo{person}{Guanting Chen}, \bibinfo{person}{Guowei Li}, \bibinfo{person}{H. Zhang}, \bibinfo{person}{Hanwei Xu}, \bibinfo{person}{Hao Yang}, \bibinfo{person}{Haowei Zhang}, \bibinfo{person}{Honghui Ding}, \bibinfo{person}{Huajian Xin}, \bibinfo{person}{Huazuo Gao}, \bibinfo{person}{Hui Li}, \bibinfo{person}{Hui Qu}, \bibinfo{person}{J.~L. Cai}, \bibinfo{person}{Jian Liang}, \bibinfo{person}{Jianzhong Guo}, \bibinfo{person}{Jiaqi Ni}, \bibinfo{person}{Jiashi Li}, \bibinfo{person}{Jin Chen}, \bibinfo{person}{Jingyang Yuan}, \bibinfo{person}{Junjie Qiu}, \bibinfo{person}{Junxiao Song}, \bibinfo{person}{Kai Dong}, \bibinfo{person}{Kaige Gao}, \bibinfo{person}{Kang Guan}, \bibinfo{person}{Lean Wang}, \bibinfo{person}{Lecong Zhang}, \bibinfo{person}{Lei Xu}, \bibinfo{person}{Leyi Xia}, \bibinfo{person}{Liang Zhao}, \bibinfo{person}{Liyue Zhang}, \bibinfo{person}{Meng Li}, \bibinfo{person}{Miaojun Wang}, \bibinfo{person}{Mingchuan Zhang}, \bibinfo{person}{Minghua Zhang}, \bibinfo{person}{Minghui Tang}, \bibinfo{person}{Mingming Li}, \bibinfo{person}{Ning Tian}, \bibinfo{person}{Panpan Huang}, \bibinfo{person}{Peiyi Wang}, \bibinfo{person}{Peng Zhang}, \bibinfo{person}{Qihao Zhu}, \bibinfo{person}{Qinyu Chen}, \bibinfo{person}{Qiushi Du}, \bibinfo{person}{R.~J. Chen}, \bibinfo{person}{R.~L. Jin}, \bibinfo{person}{Ruiqi Ge}, \bibinfo{person}{Ruizhe Pan}, \bibinfo{person}{Runxin Xu}, \bibinfo{person}{Ruyi Chen}, \bibinfo{person}{S.~S. Li}, \bibinfo{person}{Shanghao Lu}, \bibinfo{person}{Shangyan Zhou}, \bibinfo{person}{Shanhuang Chen}, \bibinfo{person}{Shaoqing Wu}, \bibinfo{person}{Shengfeng Ye}, \bibinfo{person}{Shirong Ma}, \bibinfo{person}{Shiyu Wang}, \bibinfo{person}{Shuang Zhou}, \bibinfo{person}{Shuiping Yu}, \bibinfo{person}{Shunfeng Zhou}, \bibinfo{person}{Size Zheng}, \bibinfo{person}{T. Wang}, \bibinfo{person}{Tian Pei}, \bibinfo{person}{Tian Yuan}, \bibinfo{person}{Tianyu Sun}, \bibinfo{person}{W.~L. Xiao}, \bibinfo{person}{Wangding Zeng}, \bibinfo{person}{Wei An}, \bibinfo{person}{Wen Liu}, \bibinfo{person}{Wenfeng Liang}, \bibinfo{person}{Wenjun Gao}, \bibinfo{person}{Wentao Zhang}, \bibinfo{person}{X.~Q. Li}, \bibinfo{person}{Xiangyue Jin}, \bibinfo{person}{Xianzu Wang}, \bibinfo{person}{Xiao Bi}, \bibinfo{person}{Xiaodong Liu}, \bibinfo{person}{Xiaohan Wang}, \bibinfo{person}{Xiaojin Shen}, \bibinfo{person}{Xiaokang Chen}, \bibinfo{person}{Xiaosha Chen}, \bibinfo{person}{Xiaotao Nie}, \bibinfo{person}{Xiaowen Sun}, \bibinfo{person}{Xiaoxiang Wang}, \bibinfo{person}{Xin Liu}, \bibinfo{person}{Xin Xie}, \bibinfo{person}{Xingkai Yu}, \bibinfo{person}{Xinnan Song}, \bibinfo{person}{Xinyi Zhou}, \bibinfo{person}{Xinyu Yang}, \bibinfo{person}{Xuan Lu}, \bibinfo{person}{Xuecheng Su}, \bibinfo{person}{Y. Wu}, \bibinfo{person}{Y.~K. Li}, \bibinfo{person}{Y.~X. Wei}, \bibinfo{person}{Y.~X. Zhu}, \bibinfo{person}{Yanhong Xu}, \bibinfo{person}{Yanping Huang}, \bibinfo{person}{Yao Li}, \bibinfo{person}{Yao Zhao}, \bibinfo{person}{Yaofeng Sun}, \bibinfo{person}{Yaohui Li}, \bibinfo{person}{Yaohui Wang}, \bibinfo{person}{Yi Zheng}, \bibinfo{person}{Yichao Zhang}, \bibinfo{person}{Yiliang Xiong}, \bibinfo{person}{Yilong Zhao}, \bibinfo{person}{Ying He}, \bibinfo{person}{Ying Tang}, \bibinfo{person}{Yishi Piao}, \bibinfo{person}{Yixin Dong}, \bibinfo{person}{Yixuan Tan}, \bibinfo{person}{Yiyuan Liu}, \bibinfo{person}{Yongji Wang}, \bibinfo{person}{Yongqiang Guo}, \bibinfo{person}{Yuchen Zhu}, \bibinfo{person}{Yuduan Wang}, \bibinfo{person}{Yuheng Zou}, \bibinfo{person}{Yukun Zha}, \bibinfo{person}{Yunxian Ma}, \bibinfo{person}{Yuting Yan}, \bibinfo{person}{Yuxiang You}, \bibinfo{person}{Yuxuan Liu}, \bibinfo{person}{Z.~Z. Ren}, \bibinfo{person}{Zehui Ren}, \bibinfo{person}{Zhangli Sha}, \bibinfo{person}{Zhe Fu}, \bibinfo{person}{Zhen Huang}, \bibinfo{person}{Zhen Zhang}, \bibinfo{person}{Zhenda Xie}, \bibinfo{person}{Zhewen Hao}, \bibinfo{person}{Zhihong Shao}, \bibinfo{person}{Zhiniu Wen}, \bibinfo{person}{Zhipeng Xu}, \bibinfo{person}{Zhongyu Zhang}, \bibinfo{person}{Zhuoshu Li}, \bibinfo{person}{Zihan Wang}, \bibinfo{person}{Zihui Gu}, \bibinfo{person}{Zilin Li}, {and} \bibinfo{person}{Ziwei Xie}.} \bibinfo{year}{2024}\natexlab{a}.
\newblock \bibinfo{title}{DeepSeek-V2: A Strong, Economical, and Efficient Mixture-of-Experts Language Model}.
\newblock
\newblock
\showeprint[arxiv]{2405.04434}~[cs.CL]
\urldef\tempurl%
\url{https://arxiv.org/abs/2405.04434}
\showURL{%
\tempurl}


\bibitem[DeepSeek-AI et~al\mbox{.}(2024b)]%
        {deepseekai2024deepseekv3technicalreport}
\bibfield{author}{\bibinfo{person}{DeepSeek-AI}, \bibinfo{person}{Aixin Liu}, \bibinfo{person}{Bei Feng}, \bibinfo{person}{Bing Xue}, \bibinfo{person}{Bingxuan Wang}, \bibinfo{person}{Bochao Wu}, \bibinfo{person}{Chengda Lu}, \bibinfo{person}{Chenggang Zhao}, \bibinfo{person}{Chengqi Deng}, \bibinfo{person}{Chenyu Zhang}, \bibinfo{person}{Chong Ruan}, \bibinfo{person}{Damai Dai}, \bibinfo{person}{Daya Guo}, \bibinfo{person}{Dejian Yang}, \bibinfo{person}{Deli Chen}, \bibinfo{person}{Dongjie Ji}, \bibinfo{person}{Erhang Li}, \bibinfo{person}{Fangyun Lin}, \bibinfo{person}{Fucong Dai}, \bibinfo{person}{Fuli Luo}, \bibinfo{person}{Guangbo Hao}, \bibinfo{person}{Guanting Chen}, \bibinfo{person}{Guowei Li}, \bibinfo{person}{H. Zhang}, \bibinfo{person}{Han Bao}, \bibinfo{person}{Hanwei Xu}, \bibinfo{person}{Haocheng Wang}, \bibinfo{person}{Haowei Zhang}, \bibinfo{person}{Honghui Ding}, \bibinfo{person}{Huajian Xin}, \bibinfo{person}{Huazuo Gao}, \bibinfo{person}{Hui Li}, \bibinfo{person}{Hui Qu}, \bibinfo{person}{J.~L. Cai}, \bibinfo{person}{Jian Liang}, \bibinfo{person}{Jianzhong Guo}, \bibinfo{person}{Jiaqi Ni}, \bibinfo{person}{Jiashi Li}, \bibinfo{person}{Jiawei Wang}, \bibinfo{person}{Jin Chen}, \bibinfo{person}{Jingchang Chen}, \bibinfo{person}{Jingyang Yuan}, \bibinfo{person}{Junjie Qiu}, \bibinfo{person}{Junlong Li}, \bibinfo{person}{Junxiao Song}, \bibinfo{person}{Kai Dong}, \bibinfo{person}{Kai Hu}, \bibinfo{person}{Kaige Gao}, \bibinfo{person}{Kang Guan}, \bibinfo{person}{Kexin Huang}, \bibinfo{person}{Kuai Yu}, \bibinfo{person}{Lean Wang}, \bibinfo{person}{Lecong Zhang}, \bibinfo{person}{Lei Xu}, \bibinfo{person}{Leyi Xia}, \bibinfo{person}{Liang Zhao}, \bibinfo{person}{Litong Wang}, \bibinfo{person}{Liyue Zhang}, \bibinfo{person}{Meng Li}, \bibinfo{person}{Miaojun Wang}, \bibinfo{person}{Mingchuan Zhang}, \bibinfo{person}{Minghua Zhang}, \bibinfo{person}{Minghui Tang}, \bibinfo{person}{Mingming Li}, \bibinfo{person}{Ning Tian}, \bibinfo{person}{Panpan Huang}, \bibinfo{person}{Peiyi Wang}, \bibinfo{person}{Peng Zhang}, \bibinfo{person}{Qiancheng Wang}, \bibinfo{person}{Qihao Zhu}, \bibinfo{person}{Qinyu Chen}, \bibinfo{person}{Qiushi Du}, \bibinfo{person}{R.~J. Chen}, \bibinfo{person}{R.~L. Jin}, \bibinfo{person}{Ruiqi Ge}, \bibinfo{person}{Ruisong Zhang}, \bibinfo{person}{Ruizhe Pan}, \bibinfo{person}{Runji Wang}, \bibinfo{person}{Runxin Xu}, \bibinfo{person}{Ruoyu Zhang}, \bibinfo{person}{Ruyi Chen}, \bibinfo{person}{S.~S. Li}, \bibinfo{person}{Shanghao Lu}, \bibinfo{person}{Shangyan Zhou}, \bibinfo{person}{Shanhuang Chen}, \bibinfo{person}{Shaoqing Wu}, \bibinfo{person}{Shengfeng Ye}, \bibinfo{person}{Shengfeng Ye}, \bibinfo{person}{Shirong Ma}, \bibinfo{person}{Shiyu Wang}, \bibinfo{person}{Shuang Zhou}, \bibinfo{person}{Shuiping Yu}, \bibinfo{person}{Shunfeng Zhou}, \bibinfo{person}{Shuting Pan}, \bibinfo{person}{T. Wang}, \bibinfo{person}{Tao Yun}, \bibinfo{person}{Tian Pei}, \bibinfo{person}{Tianyu Sun}, \bibinfo{person}{W.~L. Xiao}, \bibinfo{person}{Wangding Zeng}, \bibinfo{person}{Wanjia Zhao}, \bibinfo{person}{Wei An}, \bibinfo{person}{Wen Liu}, \bibinfo{person}{Wenfeng Liang}, \bibinfo{person}{Wenjun Gao}, \bibinfo{person}{Wenqin Yu}, \bibinfo{person}{Wentao Zhang}, \bibinfo{person}{X.~Q. Li}, \bibinfo{person}{Xiangyue Jin}, \bibinfo{person}{Xianzu Wang}, \bibinfo{person}{Xiao Bi}, \bibinfo{person}{Xiaodong Liu}, \bibinfo{person}{Xiaohan Wang}, \bibinfo{person}{Xiaojin Shen}, \bibinfo{person}{Xiaokang Chen}, \bibinfo{person}{Xiaokang Zhang}, \bibinfo{person}{Xiaosha Chen}, \bibinfo{person}{Xiaotao Nie}, \bibinfo{person}{Xiaowen Sun}, \bibinfo{person}{Xiaoxiang Wang}, \bibinfo{person}{Xin Cheng}, \bibinfo{person}{Xin Liu}, \bibinfo{person}{Xin Xie}, \bibinfo{person}{Xingchao Liu}, \bibinfo{person}{Xingkai Yu}, \bibinfo{person}{Xinnan Song}, \bibinfo{person}{Xinxia Shan}, \bibinfo{person}{Xinyi Zhou}, \bibinfo{person}{Xinyu Yang}, \bibinfo{person}{Xinyuan Li}, \bibinfo{person}{Xuecheng Su}, \bibinfo{person}{Xuheng Lin}, \bibinfo{person}{Y.~K. Li}, \bibinfo{person}{Y.~Q. Wang}, \bibinfo{person}{Y.~X. Wei}, \bibinfo{person}{Y.~X. Zhu}, \bibinfo{person}{Yang Zhang}, \bibinfo{person}{Yanhong Xu}, \bibinfo{person}{Yanhong Xu}, \bibinfo{person}{Yanping Huang}, \bibinfo{person}{Yao Li}, \bibinfo{person}{Yao Zhao}, \bibinfo{person}{Yaofeng Sun}, \bibinfo{person}{Yaohui Li}, \bibinfo{person}{Yaohui Wang}, \bibinfo{person}{Yi Yu}, \bibinfo{person}{Yi Zheng}, \bibinfo{person}{Yichao Zhang}, \bibinfo{person}{Yifan Shi}, \bibinfo{person}{Yiliang Xiong}, \bibinfo{person}{Ying He}, \bibinfo{person}{Ying Tang}, \bibinfo{person}{Yishi Piao}, \bibinfo{person}{Yisong Wang}, \bibinfo{person}{Yixuan Tan}, \bibinfo{person}{Yiyang Ma}, \bibinfo{person}{Yiyuan Liu}, \bibinfo{person}{Yongqiang Guo}, \bibinfo{person}{Yu Wu}, \bibinfo{person}{Yuan Ou}, \bibinfo{person}{Yuchen Zhu}, \bibinfo{person}{Yuduan Wang}, \bibinfo{person}{Yue Gong}, \bibinfo{person}{Yuheng Zou}, \bibinfo{person}{Yujia He}, \bibinfo{person}{Yukun Zha}, \bibinfo{person}{Yunfan Xiong}, \bibinfo{person}{Yunxian Ma}, \bibinfo{person}{Yuting Yan}, \bibinfo{person}{Yuxiang Luo}, \bibinfo{person}{Yuxiang You}, \bibinfo{person}{Yuxuan Liu}, \bibinfo{person}{Yuyang Zhou}, \bibinfo{person}{Z.~F. Wu}, \bibinfo{person}{Z.~Z. Ren}, \bibinfo{person}{Zehui Ren}, \bibinfo{person}{Zhangli Sha}, \bibinfo{person}{Zhe Fu}, \bibinfo{person}{Zhean Xu}, \bibinfo{person}{Zhen Huang}, \bibinfo{person}{Zhen Zhang}, \bibinfo{person}{Zhenda Xie}, \bibinfo{person}{Zhengyan Zhang}, \bibinfo{person}{Zhewen Hao}, \bibinfo{person}{Zhibin Gou}, \bibinfo{person}{Zhicheng Ma}, \bibinfo{person}{Zhigang Yan}, \bibinfo{person}{Zhihong Shao}, \bibinfo{person}{Zhipeng Xu}, \bibinfo{person}{Zhiyu Wu}, \bibinfo{person}{Zhongyu Zhang}, \bibinfo{person}{Zhuoshu Li}, \bibinfo{person}{Zihui Gu}, \bibinfo{person}{Zijia Zhu}, \bibinfo{person}{Zijun Liu}, \bibinfo{person}{Zilin Li}, \bibinfo{person}{Ziwei Xie}, \bibinfo{person}{Ziyang Song}, \bibinfo{person}{Ziyi Gao}, {and} \bibinfo{person}{Zizheng Pan}.} \bibinfo{year}{2024}\natexlab{b}.
\newblock \bibinfo{title}{DeepSeek-V3 Technical Report}.
\newblock
\newblock
\showeprint[arxiv]{2412.19437}~[cs.CL]
\urldef\tempurl%
\url{https://arxiv.org/abs/2412.19437}
\showURL{%
\tempurl}


\bibitem[Dubey et~al\mbox{.}(2024)]%
        {llama31}
\bibfield{author}{\bibinfo{person}{Abhimanyu Dubey}, \bibinfo{person}{Abhinav Jauhri}, \bibinfo{person}{Abhinav Pandey}, \bibinfo{person}{Abhishek Kadian}, \bibinfo{person}{Ahmad Al-Dahle}, \bibinfo{person}{Aiesha Letman}, \bibinfo{person}{Akhil Mathur}, \bibinfo{person}{Alan Schelten}, \bibinfo{person}{Amy Yang}, \bibinfo{person}{Angela Fan}, {et~al\mbox{.}}} \bibinfo{year}{2024}\natexlab{}.
\newblock \showarticletitle{The llama 3 herd of models}.
\newblock \bibinfo{journal}{\emph{arXiv preprint arXiv:2407.21783}} (\bibinfo{year}{2024}).
\newblock


\bibitem[Gangidi et~al\mbox{.}(2024)]%
        {gangidi2024rdma}
\bibfield{author}{\bibinfo{person}{Adithya Gangidi}, \bibinfo{person}{Rui Miao}, \bibinfo{person}{Shengbao Zheng}, \bibinfo{person}{Sai~Jayesh Bondu}, \bibinfo{person}{Guilherme Goes}, \bibinfo{person}{Hany Morsy}, \bibinfo{person}{Rohit Puri}, \bibinfo{person}{Mohammad Riftadi}, \bibinfo{person}{Ashmitha~Jeevaraj Shetty}, \bibinfo{person}{Jingyi Yang}, {et~al\mbox{.}}} \bibinfo{year}{2024}\natexlab{}.
\newblock \showarticletitle{RDMA over Ethernet for Distributed Training at Meta Scale}. In \bibinfo{booktitle}{\emph{Proceedings of the ACM SIGCOMM 2024 Conference}}. \bibinfo{pages}{57--70}.
\newblock


\bibitem[Guo et~al\mbox{.}(2024)]%
        {guo2024deepseekcoderlargelanguagemodel}
\bibfield{author}{\bibinfo{person}{Daya Guo}, \bibinfo{person}{Qihao Zhu}, \bibinfo{person}{Dejian Yang}, \bibinfo{person}{Zhenda Xie}, \bibinfo{person}{Kai Dong}, \bibinfo{person}{Wentao Zhang}, \bibinfo{person}{Guanting Chen}, \bibinfo{person}{Xiao Bi}, \bibinfo{person}{Y. Wu}, \bibinfo{person}{Y.~K. Li}, \bibinfo{person}{Fuli Luo}, \bibinfo{person}{Yingfei Xiong}, {and} \bibinfo{person}{Wenfeng Liang}.} \bibinfo{year}{2024}\natexlab{}.
\newblock \bibinfo{title}{DeepSeek-Coder: When the Large Language Model Meets Programming -- The Rise of Code Intelligence}.
\newblock
\newblock
\showeprint[arxiv]{2401.14196}~[cs.SE]
\urldef\tempurl%
\url{https://arxiv.org/abs/2401.14196}
\showURL{%
\tempurl}


\bibitem[Huang et~al\mbox{.}(2019)]%
        {huang2019gpipe}
\bibfield{author}{\bibinfo{person}{Yanping Huang}, \bibinfo{person}{Youlong Cheng}, \bibinfo{person}{Ankur Bapna}, \bibinfo{person}{Orhan Firat}, \bibinfo{person}{Dehao Chen}, \bibinfo{person}{Mia Chen}, \bibinfo{person}{HyoukJoong Lee}, \bibinfo{person}{Jiquan Ngiam}, \bibinfo{person}{Quoc~V Le}, \bibinfo{person}{Yonghui Wu}, {et~al\mbox{.}}} \bibinfo{year}{2019}\natexlab{}.
\newblock \showarticletitle{Gpipe: Efficient training of giant neural networks using pipeline parallelism}.
\newblock \bibinfo{journal}{\emph{Advances in neural information processing systems}}  \bibinfo{volume}{32} (\bibinfo{year}{2019}).
\newblock


\bibitem[Intel(2024)]%
        {gaudi3}
\bibfield{author}{\bibinfo{person}{Intel}.} \bibinfo{year}{2024}\natexlab{}.
\newblock \bibinfo{booktitle}{\emph{Intel Gaudi3}}.
\newblock
\urldef\tempurl%
\url{https://www.intel.com/content/www/us/en/products/details/processors/ai-accelerators/gaudi3.html}
\showURL{%
\tempurl}
\newblock
\shownote{Accessed: 2024.8.31}.


\bibitem[Kaplan et~al\mbox{.}(2020)]%
        {kaplan2020scaling}
\bibfield{author}{\bibinfo{person}{Jared Kaplan}, \bibinfo{person}{Sam McCandlish}, \bibinfo{person}{Tom Henighan}, \bibinfo{person}{Tom~B Brown}, \bibinfo{person}{Benjamin Chess}, \bibinfo{person}{Rewon Child}, \bibinfo{person}{Scott Gray}, \bibinfo{person}{Alec Radford}, \bibinfo{person}{Jeffrey Wu}, {and} \bibinfo{person}{Dario Amodei}.} \bibinfo{year}{2020}\natexlab{}.
\newblock \showarticletitle{Scaling laws for neural language models}.
\newblock \bibinfo{journal}{\emph{arXiv preprint arXiv:2001.08361}} (\bibinfo{year}{2020}).
\newblock


\bibitem[Kim et~al\mbox{.}(2008)]%
        {Dragonfly}
\bibfield{author}{\bibinfo{person}{John Kim}, \bibinfo{person}{Wiliam~J. Dally}, \bibinfo{person}{Steve Scott}, {and} \bibinfo{person}{Dennis Abts}.} \bibinfo{year}{2008}\natexlab{}.
\newblock \showarticletitle{Technology-Driven, Highly-Scalable Dragonfly Topology}. In \bibinfo{booktitle}{\emph{Proceedings of the 35th Annual International Symposium on Computer Architecture}} \emph{(\bibinfo{series}{ISCA '08})}. \bibinfo{publisher}{IEEE Computer Society}, \bibinfo{address}{USA}, \bibinfo{pages}{77–88}.
\newblock
\showISBNx{9780769531748}
\urldef\tempurl%
\url{https://doi.org/10.1109/ISCA.2008.19}
\showDOI{\tempurl}


\bibitem[Liu et~al\mbox{.}(2023)]%
        {liu2023ring}
\bibfield{author}{\bibinfo{person}{Hao Liu}, \bibinfo{person}{Matei Zaharia}, {and} \bibinfo{person}{Pieter Abbeel}.} \bibinfo{year}{2023}\natexlab{}.
\newblock \showarticletitle{Ring Attention with Blockwise Transformers for Near-Infinite Context}.
\newblock \bibinfo{journal}{\emph{arXiv preprint arXiv:2310.01889}} (\bibinfo{year}{2023}).
\newblock


\bibitem[Narayanan et~al\mbox{.}(2019)]%
        {narayanan2019pipedream}
\bibfield{author}{\bibinfo{person}{Deepak Narayanan}, \bibinfo{person}{Aaron Harlap}, \bibinfo{person}{Amar Phanishayee}, \bibinfo{person}{Vivek Seshadri}, \bibinfo{person}{Nikhil~R Devanur}, \bibinfo{person}{Gregory~R Ganger}, \bibinfo{person}{Phillip~B Gibbons}, {and} \bibinfo{person}{Matei Zaharia}.} \bibinfo{year}{2019}\natexlab{}.
\newblock \showarticletitle{PipeDream: Generalized pipeline parallelism for DNN training}. In \bibinfo{booktitle}{\emph{Proceedings of the 27th ACM symposium on operating systems principles}}. \bibinfo{pages}{1--15}.
\newblock


\bibitem[NVIDIA(2023)]%
        {nvidia2023dgx}
\bibfield{author}{\bibinfo{person}{NVIDIA}.} \bibinfo{year}{2023}\natexlab{}.
\newblock \bibinfo{booktitle}{\emph{DGX SuperPOD}}.
\newblock
\urldef\tempurl%
\url{https://www.nvidia.com/en-us/data-center/dgx-superpod/}
\showURL{%
\tempurl}
\newblock
\shownote{Accessed: 2024.8.31}.


\bibitem[OpenAI et~al\mbox{.}(2024)]%
        {gpt4}
\bibfield{author}{\bibinfo{person}{OpenAI}, \bibinfo{person}{Josh Achiam}, \bibinfo{person}{Steven Adler}, \bibinfo{person}{Sandhini Agarwal}, \bibinfo{person}{Lama Ahmad}, \bibinfo{person}{Ilge Akkaya}, \bibinfo{person}{Florencia~Leoni Aleman}, \bibinfo{person}{Diogo Almeida}, \bibinfo{person}{Janko Altenschmidt}, \bibinfo{person}{Sam Altman}, \bibinfo{person}{Shyamal Anadkat}, \bibinfo{person}{Red Avila}, \bibinfo{person}{Igor Babuschkin}, \bibinfo{person}{Suchir Balaji}, \bibinfo{person}{Valerie Balcom}, \bibinfo{person}{Paul Baltescu}, \bibinfo{person}{Haiming Bao}, \bibinfo{person}{Mohammad Bavarian}, \bibinfo{person}{Jeff Belgum}, \bibinfo{person}{Irwan Bello}, \bibinfo{person}{Jake Berdine}, \bibinfo{person}{Gabriel Bernadett-Shapiro}, \bibinfo{person}{Christopher Berner}, \bibinfo{person}{Lenny Bogdonoff}, \bibinfo{person}{Oleg Boiko}, \bibinfo{person}{Madelaine Boyd}, \bibinfo{person}{Anna-Luisa Brakman}, \bibinfo{person}{Greg Brockman}, \bibinfo{person}{Tim Brooks}, \bibinfo{person}{Miles Brundage}, \bibinfo{person}{Kevin Button}, \bibinfo{person}{Trevor Cai}, \bibinfo{person}{Rosie Campbell}, \bibinfo{person}{Andrew Cann}, \bibinfo{person}{Brittany Carey}, \bibinfo{person}{Chelsea Carlson}, \bibinfo{person}{Rory Carmichael}, \bibinfo{person}{Brooke Chan}, \bibinfo{person}{Che Chang}, \bibinfo{person}{Fotis Chantzis}, \bibinfo{person}{Derek Chen}, \bibinfo{person}{Sully Chen}, \bibinfo{person}{Ruby Chen}, \bibinfo{person}{Jason Chen}, \bibinfo{person}{Mark Chen}, \bibinfo{person}{Ben Chess}, \bibinfo{person}{Chester Cho}, \bibinfo{person}{Casey Chu}, \bibinfo{person}{Hyung~Won Chung}, \bibinfo{person}{Dave Cummings}, \bibinfo{person}{Jeremiah Currier}, \bibinfo{person}{Yunxing Dai}, \bibinfo{person}{Cory Decareaux}, \bibinfo{person}{Thomas Degry}, \bibinfo{person}{Noah Deutsch}, \bibinfo{person}{Damien Deville}, \bibinfo{person}{Arka Dhar}, \bibinfo{person}{David Dohan}, \bibinfo{person}{Steve Dowling}, \bibinfo{person}{Sheila Dunning}, \bibinfo{person}{Adrien Ecoffet}, \bibinfo{person}{Atty Eleti}, \bibinfo{person}{Tyna Eloundou}, \bibinfo{person}{David Farhi}, \bibinfo{person}{Liam Fedus}, \bibinfo{person}{Niko Felix}, \bibinfo{person}{Simón~Posada Fishman}, \bibinfo{person}{Juston Forte}, \bibinfo{person}{Isabella Fulford}, \bibinfo{person}{Leo Gao}, \bibinfo{person}{Elie Georges}, \bibinfo{person}{Christian Gibson}, \bibinfo{person}{Vik Goel}, \bibinfo{person}{Tarun Gogineni}, \bibinfo{person}{Gabriel Goh}, \bibinfo{person}{Rapha Gontijo-Lopes}, \bibinfo{person}{Jonathan Gordon}, \bibinfo{person}{Morgan Grafstein}, \bibinfo{person}{Scott Gray}, \bibinfo{person}{Ryan Greene}, \bibinfo{person}{Joshua Gross}, \bibinfo{person}{Shixiang~Shane Gu}, \bibinfo{person}{Yufei Guo}, \bibinfo{person}{Chris Hallacy}, \bibinfo{person}{Jesse Han}, \bibinfo{person}{Jeff Harris}, \bibinfo{person}{Yuchen He}, \bibinfo{person}{Mike Heaton}, \bibinfo{person}{Johannes Heidecke}, \bibinfo{person}{Chris Hesse}, \bibinfo{person}{Alan Hickey}, \bibinfo{person}{Wade Hickey}, \bibinfo{person}{Peter Hoeschele}, \bibinfo{person}{Brandon Houghton}, \bibinfo{person}{Kenny Hsu}, \bibinfo{person}{Shengli Hu}, \bibinfo{person}{Xin Hu}, \bibinfo{person}{Joost Huizinga}, \bibinfo{person}{Shantanu Jain}, \bibinfo{person}{Shawn Jain}, \bibinfo{person}{Joanne Jang}, \bibinfo{person}{Angela Jiang}, \bibinfo{person}{Roger Jiang}, \bibinfo{person}{Haozhun Jin}, \bibinfo{person}{Denny Jin}, \bibinfo{person}{Shino Jomoto}, \bibinfo{person}{Billie Jonn}, \bibinfo{person}{Heewoo Jun}, \bibinfo{person}{Tomer Kaftan}, \bibinfo{person}{Łukasz Kaiser}, \bibinfo{person}{Ali Kamali}, \bibinfo{person}{Ingmar Kanitscheider}, \bibinfo{person}{Nitish~Shirish Keskar}, \bibinfo{person}{Tabarak Khan}, \bibinfo{person}{Logan Kilpatrick}, \bibinfo{person}{Jong~Wook Kim}, \bibinfo{person}{Christina Kim}, \bibinfo{person}{Yongjik Kim}, \bibinfo{person}{Jan~Hendrik Kirchner}, \bibinfo{person}{Jamie Kiros}, \bibinfo{person}{Matt Knight}, \bibinfo{person}{Daniel Kokotajlo}, \bibinfo{person}{Łukasz Kondraciuk}, \bibinfo{person}{Andrew Kondrich}, \bibinfo{person}{Aris Konstantinidis}, \bibinfo{person}{Kyle Kosic}, \bibinfo{person}{Gretchen Krueger}, \bibinfo{person}{Vishal Kuo}, \bibinfo{person}{Michael Lampe}, \bibinfo{person}{Ikai Lan}, \bibinfo{person}{Teddy Lee}, \bibinfo{person}{Jan Leike}, \bibinfo{person}{Jade Leung}, \bibinfo{person}{Daniel Levy}, \bibinfo{person}{Chak~Ming Li}, \bibinfo{person}{Rachel Lim}, \bibinfo{person}{Molly Lin}, \bibinfo{person}{Stephanie Lin}, \bibinfo{person}{Mateusz Litwin}, \bibinfo{person}{Theresa Lopez}, \bibinfo{person}{Ryan Lowe}, \bibinfo{person}{Patricia Lue}, \bibinfo{person}{Anna Makanju}, \bibinfo{person}{Kim Malfacini}, \bibinfo{person}{Sam Manning}, \bibinfo{person}{Todor Markov}, \bibinfo{person}{Yaniv Markovski}, \bibinfo{person}{Bianca Martin}, \bibinfo{person}{Katie Mayer}, \bibinfo{person}{Andrew Mayne}, \bibinfo{person}{Bob McGrew}, \bibinfo{person}{Scott~Mayer McKinney}, \bibinfo{person}{Christine McLeavey}, \bibinfo{person}{Paul McMillan}, \bibinfo{person}{Jake McNeil}, \bibinfo{person}{David Medina}, \bibinfo{person}{Aalok Mehta}, \bibinfo{person}{Jacob Menick}, \bibinfo{person}{Luke Metz}, \bibinfo{person}{Andrey Mishchenko}, \bibinfo{person}{Pamela Mishkin}, \bibinfo{person}{Vinnie Monaco}, \bibinfo{person}{Evan Morikawa}, \bibinfo{person}{Daniel Mossing}, \bibinfo{person}{Tong Mu}, \bibinfo{person}{Mira Murati}, \bibinfo{person}{Oleg Murk}, \bibinfo{person}{David Mély}, \bibinfo{person}{Ashvin Nair}, \bibinfo{person}{Reiichiro Nakano}, \bibinfo{person}{Rajeev Nayak}, \bibinfo{person}{Arvind Neelakantan}, \bibinfo{person}{Richard Ngo}, \bibinfo{person}{Hyeonwoo Noh}, \bibinfo{person}{Long Ouyang}, \bibinfo{person}{Cullen O'Keefe}, \bibinfo{person}{Jakub Pachocki}, \bibinfo{person}{Alex Paino}, \bibinfo{person}{Joe Palermo}, \bibinfo{person}{Ashley Pantuliano}, \bibinfo{person}{Giambattista Parascandolo}, \bibinfo{person}{Joel Parish}, \bibinfo{person}{Emy Parparita}, \bibinfo{person}{Alex Passos}, \bibinfo{person}{Mikhail Pavlov}, \bibinfo{person}{Andrew Peng}, \bibinfo{person}{Adam Perelman}, \bibinfo{person}{Filipe de Avila Belbute~Peres}, \bibinfo{person}{Michael Petrov}, \bibinfo{person}{Henrique~Ponde de Oliveira~Pinto}, \bibinfo{person}{Michael}, \bibinfo{person}{Pokorny}, \bibinfo{person}{Michelle Pokrass}, \bibinfo{person}{Vitchyr~H. Pong}, \bibinfo{person}{Tolly Powell}, \bibinfo{person}{Alethea Power}, \bibinfo{person}{Boris Power}, \bibinfo{person}{Elizabeth Proehl}, \bibinfo{person}{Raul Puri}, \bibinfo{person}{Alec Radford}, \bibinfo{person}{Jack Rae}, \bibinfo{person}{Aditya Ramesh}, \bibinfo{person}{Cameron Raymond}, \bibinfo{person}{Francis Real}, \bibinfo{person}{Kendra Rimbach}, \bibinfo{person}{Carl Ross}, \bibinfo{person}{Bob Rotsted}, \bibinfo{person}{Henri Roussez}, \bibinfo{person}{Nick Ryder}, \bibinfo{person}{Mario Saltarelli}, \bibinfo{person}{Ted Sanders}, \bibinfo{person}{Shibani Santurkar}, \bibinfo{person}{Girish Sastry}, \bibinfo{person}{Heather Schmidt}, \bibinfo{person}{David Schnurr}, \bibinfo{person}{John Schulman}, \bibinfo{person}{Daniel Selsam}, \bibinfo{person}{Kyla Sheppard}, \bibinfo{person}{Toki Sherbakov}, \bibinfo{person}{Jessica Shieh}, \bibinfo{person}{Sarah Shoker}, \bibinfo{person}{Pranav Shyam}, \bibinfo{person}{Szymon Sidor}, \bibinfo{person}{Eric Sigler}, \bibinfo{person}{Maddie Simens}, \bibinfo{person}{Jordan Sitkin}, \bibinfo{person}{Katarina Slama}, \bibinfo{person}{Ian Sohl}, \bibinfo{person}{Benjamin Sokolowsky}, \bibinfo{person}{Yang Song}, \bibinfo{person}{Natalie Staudacher}, \bibinfo{person}{Felipe~Petroski Such}, \bibinfo{person}{Natalie Summers}, \bibinfo{person}{Ilya Sutskever}, \bibinfo{person}{Jie Tang}, \bibinfo{person}{Nikolas Tezak}, \bibinfo{person}{Madeleine~B. Thompson}, \bibinfo{person}{Phil Tillet}, \bibinfo{person}{Amin Tootoonchian}, \bibinfo{person}{Elizabeth Tseng}, \bibinfo{person}{Preston Tuggle}, \bibinfo{person}{Nick Turley}, \bibinfo{person}{Jerry Tworek}, \bibinfo{person}{Juan Felipe~Cerón Uribe}, \bibinfo{person}{Andrea Vallone}, \bibinfo{person}{Arun Vijayvergiya}, \bibinfo{person}{Chelsea Voss}, \bibinfo{person}{Carroll Wainwright}, \bibinfo{person}{Justin~Jay Wang}, \bibinfo{person}{Alvin Wang}, \bibinfo{person}{Ben Wang}, \bibinfo{person}{Jonathan Ward}, \bibinfo{person}{Jason Wei}, \bibinfo{person}{CJ Weinmann}, \bibinfo{person}{Akila Welihinda}, \bibinfo{person}{Peter Welinder}, \bibinfo{person}{Jiayi Weng}, \bibinfo{person}{Lilian Weng}, \bibinfo{person}{Matt Wiethoff}, \bibinfo{person}{Dave Willner}, \bibinfo{person}{Clemens Winter}, \bibinfo{person}{Samuel Wolrich}, \bibinfo{person}{Hannah Wong}, \bibinfo{person}{Lauren Workman}, \bibinfo{person}{Sherwin Wu}, \bibinfo{person}{Jeff Wu}, \bibinfo{person}{Michael Wu}, \bibinfo{person}{Kai Xiao}, \bibinfo{person}{Tao Xu}, \bibinfo{person}{Sarah Yoo}, \bibinfo{person}{Kevin Yu}, \bibinfo{person}{Qiming Yuan}, \bibinfo{person}{Wojciech Zaremba}, \bibinfo{person}{Rowan Zellers}, \bibinfo{person}{Chong Zhang}, \bibinfo{person}{Marvin Zhang}, \bibinfo{person}{Shengjia Zhao}, \bibinfo{person}{Tianhao Zheng}, \bibinfo{person}{Juntang Zhuang}, \bibinfo{person}{William Zhuk}, {and} \bibinfo{person}{Barret Zoph}.} \bibinfo{year}{2024}\natexlab{}.
\newblock \bibinfo{title}{GPT-4 Technical Report}.
\newblock
\newblock
\showeprint[arxiv]{2303.08774}~[cs.CL]
\urldef\tempurl%
\url{https://arxiv.org/abs/2303.08774}
\showURL{%
\tempurl}


\bibitem[Qian et~al\mbox{.}(2024)]%
        {hpn}
\bibfield{author}{\bibinfo{person}{Kun Qian}, \bibinfo{person}{Yongqing Xi}, \bibinfo{person}{Jiamin Cao}, \bibinfo{person}{Jiaqi Gao}, \bibinfo{person}{Yichi Xu}, \bibinfo{person}{Yu Guan}, \bibinfo{person}{Binzhang Fu}, \bibinfo{person}{Xuemei Shi}, \bibinfo{person}{Fangbo Zhu}, \bibinfo{person}{Rui Miao}, {et~al\mbox{.}}} \bibinfo{year}{2024}\natexlab{}.
\newblock \showarticletitle{Alibaba HPN: A Data Center Network for Large Language Model Training}.
\newblock \bibinfo{journal}{\emph{Traffic}}  \bibinfo{volume}{1} (\bibinfo{year}{2024}), \bibinfo{pages}{2}.
\newblock


\bibitem[Rajbhandari et~al\mbox{.}(2022)]%
        {ds-moe}
\bibfield{author}{\bibinfo{person}{Samyam Rajbhandari}, \bibinfo{person}{Conglong Li}, \bibinfo{person}{Zhewei Yao}, \bibinfo{person}{Minjia Zhang}, \bibinfo{person}{Reza~Yazdani Aminabadi}, \bibinfo{person}{Ammar~Ahmad Awan}, \bibinfo{person}{Jeff Rasley}, {and} \bibinfo{person}{Yuxiong He}.} \bibinfo{year}{2022}\natexlab{}.
\newblock \showarticletitle{{D}eep{S}peed-{M}o{E}: Advancing Mixture-of-Experts Inference and Training to Power Next-Generation {AI} Scale}. In \bibinfo{booktitle}{\emph{Proceedings of the 39th International Conference on Machine Learning}} \emph{(\bibinfo{series}{Proceedings of Machine Learning Research}, Vol.~\bibinfo{volume}{162})}, \bibfield{editor}{\bibinfo{person}{Kamalika Chaudhuri}, \bibinfo{person}{Stefanie Jegelka}, \bibinfo{person}{Le~Song}, \bibinfo{person}{Csaba Szepesvari}, \bibinfo{person}{Gang Niu}, {and} \bibinfo{person}{Sivan Sabato}} (Eds.). \bibinfo{publisher}{PMLR}, \bibinfo{pages}{18332--18346}.
\newblock
\urldef\tempurl%
\url{https://proceedings.mlr.press/v162/rajbhandari22a.html}
\showURL{%
\tempurl}


\bibitem[Rasley et~al\mbox{.}(2020)]%
        {deepspeed}
\bibfield{author}{\bibinfo{person}{Jeff Rasley}, \bibinfo{person}{Samyam Rajbhandari}, \bibinfo{person}{Olatunji Ruwase}, {and} \bibinfo{person}{Yuxiong He}.} \bibinfo{year}{2020}\natexlab{}.
\newblock \showarticletitle{DeepSpeed: System Optimizations Enable Training Deep Learning Models with Over 100 Billion Parameters}. In \bibinfo{booktitle}{\emph{Proceedings of the 26th ACM SIGKDD International Conference on Knowledge Discovery \& Data Mining}} (Virtual Event, CA, USA) \emph{(\bibinfo{series}{KDD '20})}. \bibinfo{publisher}{Association for Computing Machinery}, \bibinfo{address}{New York, NY, USA}, \bibinfo{pages}{3505–3506}.
\newblock
\showISBNx{9781450379984}
\urldef\tempurl%
\url{https://doi.org/10.1145/3394486.3406703}
\showDOI{\tempurl}


\bibitem[Ren et~al\mbox{.}(2023)]%
        {ren2023pangu}
\bibfield{author}{\bibinfo{person}{Xiaozhe Ren}, \bibinfo{person}{Pingyi Zhou}, \bibinfo{person}{Xinfan Meng}, \bibinfo{person}{Xinjing Huang}, \bibinfo{person}{Yadao Wang}, \bibinfo{person}{Weichao Wang}, \bibinfo{person}{Pengfei Li}, \bibinfo{person}{Xiaoda Zhang}, \bibinfo{person}{Alexander Podolskiy}, \bibinfo{person}{Grigory Arshinov}, {et~al\mbox{.}}} \bibinfo{year}{2023}\natexlab{}.
\newblock \showarticletitle{Pangu-$\{$$\backslash$Sigma$\}$: Towards trillion parameter language model with sparse heterogeneous computing}.
\newblock \bibinfo{journal}{\emph{arXiv preprint arXiv:2303.10845}} (\bibinfo{year}{2023}).
\newblock


\bibitem[SemiAnalysis(2023)]%
        {semianalysis2023gpt4}
\bibfield{author}{\bibinfo{person}{SemiAnalysis}.} \bibinfo{year}{2023}\natexlab{}.
\newblock \bibinfo{title}{GPT-4 Architecture \& Infrastructure}.
\newblock \bibinfo{howpublished}{\url{https://semianalysis.com/2023/07/10/gpt-4-architecture-infrastructure/}}.
\newblock
\newblock
\shownote{Accessed: 2025.1.9}.


\bibitem[Semianalysis(2024)]%
        {semianalysis2023}
\bibfield{author}{\bibinfo{person}{Semianalysis}.} \bibinfo{year}{2024}\natexlab{}.
\newblock \bibinfo{booktitle}{\emph{100,000 H100 Clusters: Power, Network Topology, Ethernet vs InfiniBand, Reliability, Failures, Checkpointing}}.
\newblock
\urldef\tempurl%
\url{https://www.semianalysis.com/p/100000-h100-clusters-power-network}
\showURL{%
\tempurl}


\bibitem[Shoeybi et~al\mbox{.}(2019)]%
        {shoeybi2019megatron}
\bibfield{author}{\bibinfo{person}{Mohammad Shoeybi}, \bibinfo{person}{Mostofa Patwary}, \bibinfo{person}{Raul Puri}, \bibinfo{person}{Patrick LeGresley}, \bibinfo{person}{Jared Casper}, {and} \bibinfo{person}{Bryan Catanzaro}.} \bibinfo{year}{2019}\natexlab{}.
\newblock \showarticletitle{Megatron-lm: Training multi-billion parameter language models using model parallelism}.
\newblock \bibinfo{journal}{\emph{arXiv preprint arXiv:1909.08053}} (\bibinfo{year}{2019}).
\newblock


\bibitem[Team et~al\mbox{.}(2024)]%
        {Gemma2}
\bibfield{author}{\bibinfo{person}{Gemma Team}, \bibinfo{person}{Morgane Riviere}, \bibinfo{person}{Shreya Pathak}, \bibinfo{person}{Pier~Giuseppe Sessa}, \bibinfo{person}{Cassidy Hardin}, \bibinfo{person}{Surya Bhupatiraju}, \bibinfo{person}{Léonard Hussenot}, \bibinfo{person}{Thomas Mesnard}, \bibinfo{person}{Bobak Shahriari}, \bibinfo{person}{Alexandre Ramé}, \bibinfo{person}{Johan Ferret}, \bibinfo{person}{Peter Liu}, \bibinfo{person}{Pouya Tafti}, \bibinfo{person}{Abe Friesen}, \bibinfo{person}{Michelle Casbon}, \bibinfo{person}{Sabela Ramos}, \bibinfo{person}{Ravin Kumar}, \bibinfo{person}{Charline~Le Lan}, \bibinfo{person}{Sammy Jerome}, \bibinfo{person}{Anton Tsitsulin}, \bibinfo{person}{Nino Vieillard}, \bibinfo{person}{Piotr Stanczyk}, \bibinfo{person}{Sertan Girgin}, \bibinfo{person}{Nikola Momchev}, \bibinfo{person}{Matt Hoffman}, \bibinfo{person}{Shantanu Thakoor}, \bibinfo{person}{Jean-Bastien Grill}, \bibinfo{person}{Behnam Neyshabur}, \bibinfo{person}{Olivier Bachem}, \bibinfo{person}{Alanna Walton}, \bibinfo{person}{Aliaksei Severyn}, \bibinfo{person}{Alicia Parrish}, \bibinfo{person}{Aliya Ahmad}, \bibinfo{person}{Allen Hutchison}, \bibinfo{person}{Alvin Abdagic}, \bibinfo{person}{Amanda Carl}, \bibinfo{person}{Amy Shen}, \bibinfo{person}{Andy Brock}, \bibinfo{person}{Andy Coenen}, \bibinfo{person}{Anthony Laforge}, \bibinfo{person}{Antonia Paterson}, \bibinfo{person}{Ben Bastian}, \bibinfo{person}{Bilal Piot}, \bibinfo{person}{Bo Wu}, \bibinfo{person}{Brandon Royal}, \bibinfo{person}{Charlie Chen}, \bibinfo{person}{Chintu Kumar}, \bibinfo{person}{Chris Perry}, \bibinfo{person}{Chris Welty}, \bibinfo{person}{Christopher~A. Choquette-Choo}, \bibinfo{person}{Danila Sinopalnikov}, \bibinfo{person}{David Weinberger}, \bibinfo{person}{Dimple Vijaykumar}, \bibinfo{person}{Dominika Rogozińska}, \bibinfo{person}{Dustin Herbison}, \bibinfo{person}{Elisa Bandy}, \bibinfo{person}{Emma Wang}, \bibinfo{person}{Eric Noland}, \bibinfo{person}{Erica Moreira}, \bibinfo{person}{Evan Senter}, \bibinfo{person}{Evgenii Eltyshev}, \bibinfo{person}{Francesco Visin}, \bibinfo{person}{Gabriel Rasskin}, \bibinfo{person}{Gary Wei}, \bibinfo{person}{Glenn Cameron}, \bibinfo{person}{Gus Martins}, \bibinfo{person}{Hadi Hashemi}, \bibinfo{person}{Hanna Klimczak-Plucińska}, \bibinfo{person}{Harleen Batra}, \bibinfo{person}{Harsh Dhand}, \bibinfo{person}{Ivan Nardini}, \bibinfo{person}{Jacinda Mein}, \bibinfo{person}{Jack Zhou}, \bibinfo{person}{James Svensson}, \bibinfo{person}{Jeff Stanway}, \bibinfo{person}{Jetha Chan}, \bibinfo{person}{Jin~Peng Zhou}, \bibinfo{person}{Joana Carrasqueira}, \bibinfo{person}{Joana Iljazi}, \bibinfo{person}{Jocelyn Becker}, \bibinfo{person}{Joe Fernandez}, \bibinfo{person}{Joost van Amersfoort}, \bibinfo{person}{Josh Gordon}, \bibinfo{person}{Josh Lipschultz}, \bibinfo{person}{Josh Newlan}, \bibinfo{person}{Ju yeong Ji}, \bibinfo{person}{Kareem Mohamed}, \bibinfo{person}{Kartikeya Badola}, \bibinfo{person}{Kat Black}, \bibinfo{person}{Katie Millican}, \bibinfo{person}{Keelin McDonell}, \bibinfo{person}{Kelvin Nguyen}, \bibinfo{person}{Kiranbir Sodhia}, \bibinfo{person}{Kish Greene}, \bibinfo{person}{Lars~Lowe Sjoesund}, \bibinfo{person}{Lauren Usui}, \bibinfo{person}{Laurent Sifre}, \bibinfo{person}{Lena Heuermann}, \bibinfo{person}{Leticia Lago}, \bibinfo{person}{Lilly McNealus}, \bibinfo{person}{Livio~Baldini Soares}, \bibinfo{person}{Logan Kilpatrick}, \bibinfo{person}{Lucas Dixon}, \bibinfo{person}{Luciano Martins}, \bibinfo{person}{Machel Reid}, \bibinfo{person}{Manvinder Singh}, \bibinfo{person}{Mark Iverson}, \bibinfo{person}{Martin Görner}, \bibinfo{person}{Mat Velloso}, \bibinfo{person}{Mateo Wirth}, \bibinfo{person}{Matt Davidow}, \bibinfo{person}{Matt Miller}, \bibinfo{person}{Matthew Rahtz}, \bibinfo{person}{Matthew Watson}, \bibinfo{person}{Meg Risdal}, \bibinfo{person}{Mehran Kazemi}, \bibinfo{person}{Michael Moynihan}, \bibinfo{person}{Ming Zhang}, \bibinfo{person}{Minsuk Kahng}, \bibinfo{person}{Minwoo Park}, \bibinfo{person}{Mofi Rahman}, \bibinfo{person}{Mohit Khatwani}, \bibinfo{person}{Natalie Dao}, \bibinfo{person}{Nenshad Bardoliwalla}, \bibinfo{person}{Nesh Devanathan}, \bibinfo{person}{Neta Dumai}, \bibinfo{person}{Nilay Chauhan}, \bibinfo{person}{Oscar Wahltinez}, \bibinfo{person}{Pankil Botarda}, \bibinfo{person}{Parker Barnes}, \bibinfo{person}{Paul Barham}, \bibinfo{person}{Paul Michel}, \bibinfo{person}{Pengchong Jin}, \bibinfo{person}{Petko Georgiev}, \bibinfo{person}{Phil Culliton}, \bibinfo{person}{Pradeep Kuppala}, \bibinfo{person}{Ramona Comanescu}, \bibinfo{person}{Ramona Merhej}, \bibinfo{person}{Reena Jana}, \bibinfo{person}{Reza~Ardeshir Rokni}, \bibinfo{person}{Rishabh Agarwal}, \bibinfo{person}{Ryan Mullins}, \bibinfo{person}{Samaneh Saadat}, \bibinfo{person}{Sara~Mc Carthy}, \bibinfo{person}{Sarah Cogan}, \bibinfo{person}{Sarah Perrin}, \bibinfo{person}{Sébastien M.~R. Arnold}, \bibinfo{person}{Sebastian Krause}, \bibinfo{person}{Shengyang Dai}, \bibinfo{person}{Shruti Garg}, \bibinfo{person}{Shruti Sheth}, \bibinfo{person}{Sue Ronstrom}, \bibinfo{person}{Susan Chan}, \bibinfo{person}{Timothy Jordan}, \bibinfo{person}{Ting Yu}, \bibinfo{person}{Tom Eccles}, \bibinfo{person}{Tom Hennigan}, \bibinfo{person}{Tomas Kocisky}, \bibinfo{person}{Tulsee Doshi}, \bibinfo{person}{Vihan Jain}, \bibinfo{person}{Vikas Yadav}, \bibinfo{person}{Vilobh Meshram}, \bibinfo{person}{Vishal Dharmadhikari}, \bibinfo{person}{Warren Barkley}, \bibinfo{person}{Wei Wei}, \bibinfo{person}{Wenming Ye}, \bibinfo{person}{Woohyun Han}, \bibinfo{person}{Woosuk Kwon}, \bibinfo{person}{Xiang Xu}, \bibinfo{person}{Zhe Shen}, \bibinfo{person}{Zhitao Gong}, \bibinfo{person}{Zichuan Wei}, \bibinfo{person}{Victor Cotruta}, \bibinfo{person}{Phoebe Kirk}, \bibinfo{person}{Anand Rao}, \bibinfo{person}{Minh Giang}, \bibinfo{person}{Ludovic Peran}, \bibinfo{person}{Tris Warkentin}, \bibinfo{person}{Eli Collins}, \bibinfo{person}{Joelle Barral}, \bibinfo{person}{Zoubin Ghahramani}, \bibinfo{person}{Raia Hadsell}, \bibinfo{person}{D. Sculley}, \bibinfo{person}{Jeanine Banks}, \bibinfo{person}{Anca Dragan}, \bibinfo{person}{Slav Petrov}, \bibinfo{person}{Oriol Vinyals}, \bibinfo{person}{Jeff Dean}, \bibinfo{person}{Demis Hassabis}, \bibinfo{person}{Koray Kavukcuoglu}, \bibinfo{person}{Clement Farabet}, \bibinfo{person}{Elena Buchatskaya}, \bibinfo{person}{Sebastian Borgeaud}, \bibinfo{person}{Noah Fiedel}, \bibinfo{person}{Armand Joulin}, \bibinfo{person}{Kathleen Kenealy}, \bibinfo{person}{Robert Dadashi}, {and} \bibinfo{person}{Alek Andreev}.} \bibinfo{year}{2024}\natexlab{}.
\newblock \bibinfo{title}{Gemma 2: Improving Open Language Models at a Practical Size}.
\newblock
\newblock
\showeprint[arxiv]{2408.00118}~[cs.CL]
\urldef\tempurl%
\url{https://arxiv.org/abs/2408.00118}
\showURL{%
\tempurl}


\bibitem[Touvron et~al\mbox{.}(2023)]%
        {llama}
\bibfield{author}{\bibinfo{person}{Hugo Touvron}, \bibinfo{person}{Thibaut Lavril}, \bibinfo{person}{Gautier Izacard}, \bibinfo{person}{Xavier Martinet}, \bibinfo{person}{Marie-Anne Lachaux}, \bibinfo{person}{Timothée Lacroix}, \bibinfo{person}{Baptiste Rozière}, \bibinfo{person}{Naman Goyal}, \bibinfo{person}{Eric Hambro}, \bibinfo{person}{Faisal Azhar}, \bibinfo{person}{Aurelien Rodriguez}, \bibinfo{person}{Armand Joulin}, \bibinfo{person}{Edouard Grave}, {and} \bibinfo{person}{Guillaume Lample}.} \bibinfo{year}{2023}\natexlab{}.
\newblock \bibinfo{title}{LLaMA: Open and Efficient Foundation Language Models}.
\newblock
\newblock
\showeprint[arxiv]{2302.13971}~[cs.CL]
\urldef\tempurl%
\url{https://arxiv.org/abs/2302.13971}
\showURL{%
\tempurl}


\bibitem[Wang et~al\mbox{.}({[n.\,d.]})]%
        {wangrail}
\bibfield{author}{\bibinfo{person}{Weiyang Wang}, \bibinfo{person}{Manya Ghobadi}, \bibinfo{person}{Kayvon Shakeri}, \bibinfo{person}{Ying Zhang}, {and} \bibinfo{person}{Naader Hasani}.} \bibinfo{year}{[n.\,d.]}\natexlab{}.
\newblock \showarticletitle{Rail-only: A Low-Cost High-Performance Network for Training LLMs with Trillion Parameters}.
\newblock  (\bibinfo{year}{[n.\,d.]}).
\newblock


\bibitem[WCCFtech(2024)]%
        {wccftech_2023}
\bibfield{author}{\bibinfo{person}{WCCFtech}.} \bibinfo{year}{2024}\natexlab{}.
\newblock \bibinfo{title}{Elon Musk's xAI to Use 100,000 NVIDIA H100 GPUs to Train Grok-3 AI Model, Grok-2 Launch in August}.
\newblock \bibinfo{howpublished}{\url{https://wccftech.com/elon-musk-xai-100000-nvidia-h100-gpus-train-grok-3-ai-model-grok-2-launch-august/}}.
\newblock
\newblock
\shownote{Accessed: 2024.10}.


\bibitem[xAI(2024)]%
        {xai_100k}
\bibfield{author}{\bibinfo{person}{xAI}.} \bibinfo{year}{2024}\natexlab{}.
\newblock \bibinfo{booktitle}{\emph{xAI’s 100K GPU Cluster}}.
\newblock
\urldef\tempurl%
\url{https://www.google.com/url?sa=t&rct=j&q=&esrc=s&source=web&cd=&ved=2ahUKEwiLw7q2_62JAxXo4zQHHe_ACIIQFnoECBMQAQ&url=https%3A%2F%2Fwww.perplexity.ai%2Fpage%2Fxai-brings-colossus-online-f6EQHsQ_S.egPXbSN17CmQ&usg=AOvVaw1Ln_XTwyBj-gr5OwjFFRk3&opi=89978449}
\showURL{%
\tempurl}


\bibitem[Yang et~al\mbox{.}(2023)]%
        {yang2023baichuan2openlargescale}
\bibfield{author}{\bibinfo{person}{Aiyuan Yang}, \bibinfo{person}{Bin Xiao}, \bibinfo{person}{Bingning Wang}, \bibinfo{person}{Borong Zhang}, \bibinfo{person}{Ce Bian}, \bibinfo{person}{Chao Yin}, \bibinfo{person}{Chenxu Lv}, \bibinfo{person}{Da Pan}, \bibinfo{person}{Dian Wang}, \bibinfo{person}{Dong Yan}, \bibinfo{person}{Fan Yang}, \bibinfo{person}{Fei Deng}, \bibinfo{person}{Feng Wang}, \bibinfo{person}{Feng Liu}, \bibinfo{person}{Guangwei Ai}, \bibinfo{person}{Guosheng Dong}, \bibinfo{person}{Haizhou Zhao}, \bibinfo{person}{Hang Xu}, \bibinfo{person}{Haoze Sun}, \bibinfo{person}{Hongda Zhang}, \bibinfo{person}{Hui Liu}, \bibinfo{person}{Jiaming Ji}, \bibinfo{person}{Jian Xie}, \bibinfo{person}{JunTao Dai}, \bibinfo{person}{Kun Fang}, \bibinfo{person}{Lei Su}, \bibinfo{person}{Liang Song}, \bibinfo{person}{Lifeng Liu}, \bibinfo{person}{Liyun Ru}, \bibinfo{person}{Luyao Ma}, \bibinfo{person}{Mang Wang}, \bibinfo{person}{Mickel Liu}, \bibinfo{person}{MingAn Lin}, \bibinfo{person}{Nuolan Nie}, \bibinfo{person}{Peidong Guo}, \bibinfo{person}{Ruiyang Sun}, \bibinfo{person}{Tao Zhang}, \bibinfo{person}{Tianpeng Li}, \bibinfo{person}{Tianyu Li}, \bibinfo{person}{Wei Cheng}, \bibinfo{person}{Weipeng Chen}, \bibinfo{person}{Xiangrong Zeng}, \bibinfo{person}{Xiaochuan Wang}, \bibinfo{person}{Xiaoxi Chen}, \bibinfo{person}{Xin Men}, \bibinfo{person}{Xin Yu}, \bibinfo{person}{Xuehai Pan}, \bibinfo{person}{Yanjun Shen}, \bibinfo{person}{Yiding Wang}, \bibinfo{person}{Yiyu Li}, \bibinfo{person}{Youxin Jiang}, \bibinfo{person}{Yuchen Gao}, \bibinfo{person}{Yupeng Zhang}, \bibinfo{person}{Zenan Zhou}, {and} \bibinfo{person}{Zhiying Wu}.} \bibinfo{year}{2023}\natexlab{}.
\newblock \bibinfo{title}{Baichuan 2: Open Large-scale Language Models}.
\newblock
\newblock
\showeprint[arxiv]{2309.10305}~[cs.CL]
\urldef\tempurl%
\url{https://arxiv.org/abs/2309.10305}
\showURL{%
\tempurl}


\bibitem[Zeng et~al\mbox{.}(2023)]%
        {chatglm}
\bibfield{author}{\bibinfo{person}{Aohan Zeng}, \bibinfo{person}{Xiao Liu}, \bibinfo{person}{Zhengxiao Du}, \bibinfo{person}{Zihan Wang}, \bibinfo{person}{Hanyu Lai}, \bibinfo{person}{Ming Ding}, \bibinfo{person}{Zhuoyi Yang}, \bibinfo{person}{Yifan Xu}, \bibinfo{person}{Wendi Zheng}, \bibinfo{person}{Xiao Xia}, \bibinfo{person}{Weng~Lam Tam}, \bibinfo{person}{Zixuan Ma}, \bibinfo{person}{Yufei Xue}, \bibinfo{person}{Jidong Zhai}, \bibinfo{person}{Wenguang Chen}, \bibinfo{person}{Peng Zhang}, \bibinfo{person}{Yuxiao Dong}, {and} \bibinfo{person}{Jie Tang}.} \bibinfo{year}{2023}\natexlab{}.
\newblock \bibinfo{title}{GLM-130B: An Open Bilingual Pre-trained Model}.
\newblock
\newblock
\showeprint[arxiv]{2210.02414}~[cs.CL]
\urldef\tempurl%
\url{https://arxiv.org/abs/2210.02414}
\showURL{%
\tempurl}


\end{thebibliography}
\clearpage

\end{document}